\documentclass[final,3p,times,twocolumn]{elsarticle} %%%%%%
\usepackage{amssymb}
\usepackage{amsmath}
\usepackage{hyperref}
\usepackage{ulem}
\usepackage[dvipsnames]{xcolor}
\usepackage{multirow}
\usepackage{svg}
\usepackage{graphicx}
\usepackage{amssymb}
\usepackage{lineno}

\usepackage{float}
\usepackage{epstopdf}
\usepackage{siunitx}
\usepackage{hyphenat}
\usepackage{todonotes}
\usepackage{subcaption}
\usepackage{amsmath}
\usepackage{amssymb}
\usepackage{caption}
\usepackage{graphicx}
\usepackage{latexsym}
\usepackage{times}
\usepackage{siunitx}
\usepackage[utf8]{inputenc}
\usepackage{textcomp}
\usepackage{xcolor}
\usepackage{enumitem}
\usepackage{makecell}

\usepackage[version=4]{mhchem}

\usepackage{nomencl,etoolbox,ragged2e,siunitx,mathtools}

\renewcommand\nomgroup[1]{\def\nomtemp{\csname nomstart#1\endcsname}\nomtemp}

\makenomenclature

\DeclareSIUnit\volper{vol\%}
\DeclareSIUnit\massper{\%_{m}}
\usepackage{array}
\DeclareMathAlphabet{\mathpzc}{OT1}{pzc}{m}{it}
\newcolumntype{P}[1]{>{\centering\arraybackslash}p{#1}}
\hypersetup{
colorlinks=true,
linkcolor=blue,
citecolor=blue,
citebordercolor={0 1 0}
}
\journal{Fuel}

\begin{document}

\begin{frontmatter}

\title{Critical nanoparticle formation in iron combustion: single particle experiments with \textit{in-situ} multi-parameter diagnostics aided by multi-scale simulations}

% \title{Critical single-particle iron combustion in hot oxidizing flows: Part II: Microexplosion}



\author[rsm]{Tao Li\corref{cor1}}
\ead{tao.li@rsm.tu-darmstadt.de}
\author[stfs]{Bich-Diep Nguyen}
\author[pen]{Yawei Gao}
\author[rsm]{Daoguan Ning}
\author[rsm]{Benjamin Böhm}
\author[stfs]{Arne Scholtissek}
\author[pen]{Adri C.T. van Duin}
\author[stfs]{Christian Hasse}
\author[rsm]{Andreas Dreizler}

\cortext[cor1]{Corresponding author}

\address[rsm]{Technical University of Darmstadt, Department of Mechanical Engineering, Reactive Flows and Diagnostics, Otto-Berndt Straße 3, 64287 Darmstadt, Germany}
\address[stfs]{Technical University of Darmstadt, Mechanical Engineering, Institute for Simulation of reactive Thermo-Fluid Systems, Otto-Berndt Straße 2, 64287 Darmstadt, Germany}
\address[pen]{Department of Mechanical Engineering, Pennsylvania State University, University Park, 16802, PA, United States}

\begin{abstract}
The formation of iron oxide nanoparticle (NP) presents challenges such as efficiency penalties and fine dust emissions in practical iron combustion systems, necessitating a deeper academic understanding of the underlying formation mechanisms and critical thermochemical conditions.~This study, utilizing both experiments and multi-scale simulation tools, investigates the NP clouds generated by single iron particles burning in high-temperature oxidizing environments.~The ambient gas conditions were provided by a laminar flat flame burner, where the post-flame oxygen mole fraction was varied between 20, 30, and 40 vol\%, with a constant gas temperature of approximately 1800\,K.~In the experiments, high-speed \textit{in-situ} diagnostics were employed to simultaneously measure particle size, NP initiation, NP cloud evolution, and the surface temperature history of the microparticles.~The setup involved three 10\,kHz imaging systems: one for two-color pyrometry and two for diffusive-backlight illumination (DBI), specifically targeting particle size and nanoparticle measurements.~The study demonstrates the powerful capabilities of these multi-physics diagnostics, allowing for precise quantification of NP initiation time and temperature, which were found to depend on both particle size and ambient oxygen concentration.~Detailed CFD simulations revealed that the enhanced convection velocity, driven by increased Stefan flow, transports NPs toward the parent iron particles, particularly under high-oxygen conditions.~This phenomenon delayed the appearance of detectable NP clouds, leading to higher apparent microparticle temperatures at the point of NP initiation.~This insight complemented the experimental findings, providing a comprehensive explanation of the observed NP-cloud-initiation temperature that increases with higher oxygen levels.~Further analysis through molecular dynamics (MD) simulations uncovered the role of \ce{FeO2}(g) as a key NP precursor, which forms when \ce{Fe} atoms were dissociated from the liquid phase.~The simulations also showed that the initial temperature significantly influenced the chemical composition of the resulting nanoclusters, with Fe(II) predominating at higher temperatures and Fe(III) at lower temperatures.~This integrated approach not only advances our understanding of NP formation in iron combustion but also provides valuable insights into the conditions that affect nanoparticle characteristics.
\end{abstract}

\begin{keyword}
Iron powder \sep Nanoparticle formation \sep Single particle combustion \sep Multi-parameter in-situ diagnostics \sep Multi-scale simulations
\end{keyword}

\end{frontmatter}

\section{Introduction}
\label{sec:intro}

\subsection{Iron as recyclable energy carrier}
\label{subsec:intro1}

% Why iron: advantages and promises
In 2023, renewable energy sources set a new record, accounting for 30\% of global electricity generation \cite{IEA.2024}.~In Germany, this figure reached 40\% in 2022, showing continued robust growth with no signs of slowing down \cite{IEA.2022}.~As the price of photovoltaic (PV) technology has significantly decreased in recent years, the primary cost of renewable electricity now lies in energy storage rather than generation or transmission.~Energy storage is essential for mitigating the intermittency of renewable electricity production, which varies greatly depending on geography and time.~For long-term energy storage capacity, non-carbonaceous chemical energy carriers are mandatory, as batteries remains prohibitively expensive in terms of the cost per unit of power.~Among these, hydrogen (\ce{H2}), ammonia (\ce{NH3}), and iron (Fe) are particularly interesting chemicals, allowing for low-emission energy conversion and recyclable utilization.

Iron offers unique advantages over other energy storage candidates, considering that it has the potential to replace coal, providing a cost-effective solution for simultaneously retrofitting and decarbonizing existing power plants \cite{Neumann.2024}.~In this concept, thermal energy can be released by directly combusting iron powders with air in dust-firing or fluidized-bed processes \cite{Bergthorson.2015.Appl.Energy}, producing heat and electricity.~The combustion products, primarily micron-sized iron oxides, can be collected from the exhaust gas and chemically reduced through electrolysis using renewable electricity, or through thermochemical reduction utilizing green hydrogen \cite{Bergthorson.2018.Prog.EnergyCombust.Sci., Neumann.2024}.,~This so-called combined reduction-oxidation (redox) concept, targeting its specific application scenarios, demonstrates high potential for energy decarbonization and energy security \cite{Dreizler.2021.ApplEnergyCombustSci}.~Consequently, there is growing academic interest in fundamental studies of the iron combustion, such as single iron particle oxidation processes, including modeling \cite{Mi.2022,Fujinawa.2023,Nguyen.2024,Mich.2024} and experiments \cite{Ning.2022.Combust.Flame, Panahi.2022.ApplEnergyCombustSci, Li.2022.Combust.Flameb,Ning.2023, Hameete.2024}, as well as flame dynamics in laminar and turbulent flows \cite{Tang.2011,McRae.2019, Fedoryk.2023.ApplEnergyCombustSci, Krenn.2024, Goroshin.2022.Prog.EnergyCombust.Sci.}, to name a few.

% Challenge: among many, nanoparticle formation is one of the biggest issue: fein particle emission and mass loss.
One of the unresolved challenges in iron combustion is the undesired formation of nanoparticles (NPs).~Traditionally, iron oxidation has been considered a heterogeneous process, as the flame temperature is predicted to be below the boiling point \cite{Bergthorson.2015.Appl.Energy} and heterogeneous surface reactions prevail.~This heterogeneous nature was regarded as a key for the recyclability of iron as a chemical fuel.~However, numerous experiments have provided clear evidence of NP formation \cite{Li.2021.Opt.Express, Li.2022.Combust.Flame, Ning.2022.Combust.Flame, Krenn.2024, Cen.2024}.~These experiments underscore the urgent need to better understand the mechanisms, especially the thermochemical conditions in favor of NP formation.~Such understanding is crucial for guiding practical applications, where modifying thermal, chemical, and aerodynamic conditions could help to minimize mass loss and the emission of nano oxides into the atmosphere.~With this motivation, the current study aims to provide new insights into key parameters and processes through a predominant experimental investigation aided by multi-scale simulations.

\subsection{Reviewing single-particle experiments}
\label{subsec:intro2}

% What missions should single-particle studies fulfill?: fundamental understanding and model validation
% What contributions have the in-situ diagnostics? Size, luminosity/oxidation stage, NP, temperature.

Across various research approaches and physical scales, single iron particle experiments are critical for advancing our understanding of oxidation reactions and their interactions with surrounding gases and flows.~Recent advancements in \textit{in-situ} laser diagnostics have significantly broadened the range of parameters that can be experimentally accessed \cite{Alden.2022.Proc.Combust.Inst., Li.2024}.~Enabled by these diagnostics, previous experimental studies have delved into progressive oxidation stages \cite{Ning.2022.Combust.Flame, Ning.2023}, ignition phenomena \cite{Ning.2024, Abdallah.2024}, particle temperature histories \cite{Panahi.2022.ApplEnergyCombustSci, Ning.2022.Combust.Flameb, Hameete.2024}, nanoparticle formation \cite{Li.2022.Combust.Flameb, Li.2022.Combust.Flame, Ning.2022.Proc.Combust.Inst.}, and microexplosions \cite{Li.2021.Opt.Express, Huang.2022.Combust.Flame}.~These datasets are essential for describing the underlying physico-chemical processes and play a crucial role in validating and refining mathematical models.~Efforts to integrate experimental data into numerical simulations have been demonstrated in several studies, including those by \cite{Fujinawa.2023, Nguyen.2024, Mich.2024, Thijs.2023, Thijs.2024}.

% What is the state-of-the-art on nanoparticle formation, including experiments
So far, only a few optical experiments were dedicated to \textit{in-situ} measurements of NP formation, as reviewed in \cite{Li.2023}.~For instance, Li et al.~\cite{Li.2021.Opt.Express} observed the release of nanoparticle clouds from 70\,\textmu m iron particles burning in the hot exhaust gases of a flat flame stabilized on a McKenna burner, using time-resolved shadowgraphy and thermal imaging.~They also reported spectrally resolved emission lines vapor-phase FeO*, which indicates the existing precursors for NPs.~Ning et al.~\cite{Ning.2022.Proc.Combust.Inst.} measured laser-ignited iron particles burning in room-temperature laminar flows with varying oxygen concentrations, revealing the correlation between NPs' evolution and parent particle temperature.~More recently, Cen et al.~\cite{Cen.2024} reported the volume fraction of NP clouds using the light attenuation method.~Although uncertainties remain in quantifying the conversion rate due to the unknown optical properties of NPs, there is clear evidence showing an increase in NP concentrations at higher parent particle temperatures.

In addition to these \textit{in-situ} measurements, numerous \textit{ex-situ} analyses of collected iron combustion products have confirmed significant NP formation, such as in~\cite{Ning.2022.Combust.Flame, Toth.2020.PowderTechnol.,  Poletaev.2020.Combust.Sci.Technol.}.

\subsection{Scope of the current study}
\label{subsec:intro3}

Despite previous efforts, the formation of nano iron oxides during high-temperature oxidation remains a largely under-explored research area with many unanswered questions.~In our previous study \cite{Li.2022.Combust.Flameb}, we employed simultaneous diffuse backlight illumination (DBI) and luminosity (LU) imaging to temporally visualize the initiation and growth of nanoparticle clouds.~We found that the temporal evolution of these clouds was dependent on particle size and correlated with surface temperature, as qualitatively indicated by thermal radiation intensity.~However, several key questions remain, including the underlying mechanism of nanoparticle formation, the initial microparticle temperature at the onset of NP formation, and the quantitative influence of particle size and ambient oxygen levels.~These unresolved issues have driven the current study to seek further insights.

To obtain more detailed experimental data, this study enhances the previous diagnostic approach by incorporating \textit{in-situ} particle sizing and surface temperature measurements.~We utilize simultaneous multi-physics optical diagnostics, including DBI sizing, DBI nano, and two-color pyrometry, all with a temporal resolution of 10\,kHz.~This advanced diagnostic setup enables the quantification of multiple parameters, such as initial particle size, nanoparticle initiation time, and the corresponding parent particle temperatures.~Additionally, simulations are conducted at various scales to complement the experimental observations and interpretations.~Detailed CFD simulations and molecular dynamics (MD) simulations, focused on particle and molecular levels respectively, are performed to provide insights into the processes of nanoparticle cloud formation and detachment.~Through this integrated approach, the study also aims to establish a comprehensive dataset for validating nanoparticle modeling.

The experimental and numerical methodologies are outlined in Section \ref{sec:experiments} and Section \ref{sec:simulation tools}, respectively.~The results are presented and discussed in Section \ref{sec:R&D}, with a focus on parameter trajectories in Section \ref{subsec:parametertrajectories}, surface temperatures in Section \ref{subsec:particletemperature}, NP cloud initiation in Section \ref{subsec:NP initiation}, NP cloud growth in Section \ref{subsec:NP growth}, and precursor formation and agglomeration in Section \ref{subsec:precursor formation}.~Conclusions are summarized in Section \ref{sec:conclusions}.

\section{Experiments, Materials, and Diagnostics}
\label{sec:experiments}

\subsection{Flow reactor and combustion atmospheres}
\label{subsec:reactor}

\begin{figure*}[h!]
\centering
\includegraphics[width=140mm]{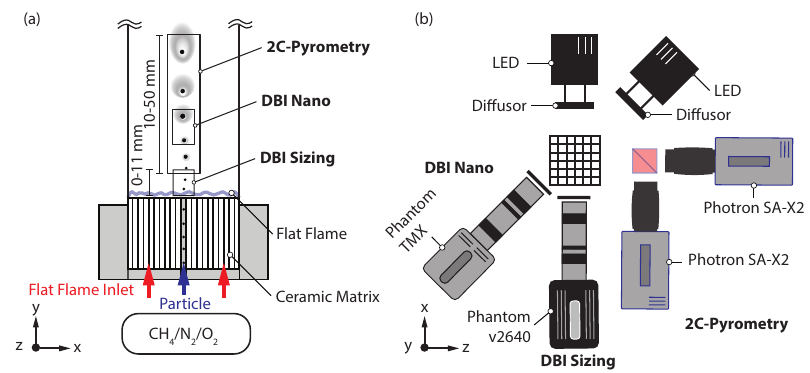}
    \caption{(a) Schematic drawing of the laminar flow reactor and field of views of \textit{in-situ} optical diagnostics.~(b) Sketch of the optical layout of simultaneous particle sizing (DBI sizing), nanoparticle formation (DBI Nano), and particle surface temperature measurements (2C-Pyrometry).}
\label{Fig: Exp Setup}
\end{figure*}

To isolate influential parameters and facilitate validation of numerical simulations, iron oxidation should occur in a controlled gas atmosphere.~For this purpose, combustion experiments of single iron particles were performed in a well-characterized laminar flow reactor (LFR), as depicted in Fig.~\ref{Fig: Exp Setup}(a).~This reactor provides defined thermal, chemical, and aerodynamic boundary conditions.~The LFR consisted of a 80\,$\times$\,\SI{80}{\milli\meter} ceramic honeycomb matrix, a rectangular fused silica enclosure ensuring optical access, and a particle seeding unit connected to a central injection tube with an inner diameter of \SI{0.8}{\milli\meter}.~Lean premixed methane flames were stabilized on the top of the ceramic surface with bulk flow velocities matching the laminar flame speeds, ensuring a flat flame surface.~The axial flame location is about 1.5\,mm above the burner surface, determining the particle heating initiation, which was measured by laser-induced fluorescence of OH radicals (OH-LIF) \cite{Li.2021.Fuel}.~In this study, three different inlet mixtures of \ce{CH4}/\ce{N2}/\ce{O2} were used, producing excess \ce{O2} with mole fractions of 0.2, 0.3, and 0.4 in the post-flame flows.~Accordingly, these conditions are denoted as AIR20, AIR30, and AIR40, respectively, which are detailed in Table~\ref{T:Condition} showing mole fractions of major molecules and mean gas temperature $T_\textrm{m}$.~The gas compositions are obtained from chemical equilibrium calculations using the GRI3.0 mechanism in Cantera \cite{Goodwin.2021}, which considers the volumetric ratio of inlet gases given in \cite{Li.2021.Fuel}.~ The gas temperature $T_\textrm{m}$ is averaged within 0\,--\,50\,mm height above burner (HAB) along the burner central temperature profile previously measured using OH-LIF combined with laser absorption spectroscopy \cite{Li.2021.Fuel}.~The accuracy of the temperature was validated against the measurements results of nanosecond rotational-vibrational coherent anti-Stokes Raman spectroscopy (CARS) of \ce{N2} \cite{Koser.2019.Proc.Combust.Inst.}, showing quantitative agreements.~The gas velocities were measured using particle image velocimetry (PIV) without fuel particles, detailed in \cite{Li.2021.Fuel}, showing a rapid increase when passing the flat flame and a plateau velocity of about 1.67\,m/s for all investigated conditions.~In summary, all three conditions have similar thermal, chemical, and aerodynamic boundary conditions, allowing for exploring the isolated effects of oxygen content on iron oxidation processes.

% Fig: Experimental setup

\begin{table}[!htb]
\centering
\small
\caption{Gas composition and axial mean temperature in the laminar flow reactor.}
\begin{tabular}{cccccc}
\hline
Conditions & $X_\textrm{\ce{O_2}}$ & $X_\textrm{\ce{N2}}$ & $X_\textrm{\ce{CO2}}$ & $X_\textrm{\ce{H2O}}$ & $T_\textrm{m}$ \\ \hline
AIR20      & 0.20          &  0.596     &     0.067   &    0.135    &  1753 \\
AIR30      & 0.30          &   0.497    &    0.067    &    0.135    &  1739 \\ 
AIR40      & 0.40          &   0.412    &     0.070   &     0.141    & 1739 \\ \hline
\end{tabular}
\label{T:Condition}
\end{table}

\subsection{Multi-physics optical diagnostics}
\label{subsec:diagnostics}

Single iron particle combustion is a complex multi-phase (i.e., solid, liquid and vapor) and multi-physics (convection, diffusion, oxidation, and radiation) phenomenon.~Unfortunately, \textit{in-situ} optical diagnostics, which have been largely advanced for gas-phase reactions, are still limited for such a predominantly heterogeneous process \cite{Li.2023}.~In this work, we attempted to use multi-parameter, \textit{in-situ} optical diagnostics to measure particle size, formation of nanometric iron oxide clouds, and surface temperature simultaneously for each individual iron particle injected into the LFR.~Figure~\ref{Fig: Exp Setup}(b) illustrates the experimental setup, including a two-color pyrometer (2C-Pyrometry) and two diffuse backlight-illumination (DBI) systems, namely DBI nano and DBI sizing to measure particle size and nanoparticle cloud, respectively.~The arrangement of region of interests (ROI) is illustrated in Fig.~\ref{Fig: Exp Setup}(a).~Three diagnostic systems were synchronized at a recording rate of 10\,kHz, enabling temporal correlation of multiple parameters along the oxidation progress of each particle. 

% DBI nano
The size of iron particles during oxidation undergoes a dynamic change due to the variation in bulk density and composition \cite{Ning.2022.Proc.Combust.Inst., Li.2022.Combust.Flameb}, thus size-resolved measurements are crucial to reduce the experimental uncertainty due to size variations.~In this study, the initial particle size needs to be \textit{in-situ} determined to perform conditional analysis of the particle temperature.~As the particle enters the laminar flow reactor, it was imaged between 0\,--\,11\,mm HAB by the DBI sizing system, which was established to resolve the morphology of micrometer-sized solid fuel particles \cite{Li.2021.RenewableSustainableEnergyRev., Li.2022.Proc.Combust.Inst.b}.~Backwardly illuminated by a high-power pulsed LED (LPS3, ILA, 10\,\textmu s pulse width, 10\,kHz), particle shadow images were captured by a high-speed CMOS camera (v2640, Phantom).~The 10\,\textmu s illumination pulse was sufficiently short to freeze the particle motion, while maximizing the signal-to-background ratio.~The signal intensity was further optimized by suppressing the gas and solid luminescence using a bandpass filter (525\,$\pm$\,25\,nm), matching the spectral peak of the LED light at approximately 518\,nm.~To resolve micrometric iron particles spatially, the high-speed camera was coupled with a long-distance microscope (Infinity K2, CF-1/B lens).~The projected pixel size was 5.2\,\textmu m calibrated by using a high-precision target (Thorlabs, R2L2S3P1).

% Pyrometry
As one of the most relevant parameters, particle surface temperature is of importance for understanding the oxidation stages \cite{Ning.2022.Combust.Flameb} and validating particle combustion models \cite{Mich.2024}.~The surface temperature can be determined using pyrometry by assuming that the burning iron particle is a gray-body emitter over a broad temperature range \cite{Ning.2024, Hameete.2024}.~In this study, when particles were within 10\,--\,50\,mm HAB, surface temperatures were measured by a customized 2C pyrometer, as initially introduced in \cite{Ning.2024}.~As illustrated in Fig.~\ref{Fig: Exp Setup}(b), a 50/50 beam splitter cube (Thorlabs BS032) divided thermal radiation equally to two identical high-speed CMOS cameras (Photron SA-X2).~The cameras were equipped with bandpass filters with high transmission at either 850$\pm$25\,nm or 950$\pm$25\,nm.~An additional ND filter (ND$=$0.6) was used for the 850\,nm-camera, leveraging a similar intensity level of both cameras to improve the measurement dynamic range.~Two identical camera lenses (Sigma Macro 105 $f$2.8) were used with a fully open aperture.~The projected pixel size was about 42\,\textmu m/pixel with a FOV marked in  Fig.~\ref{Fig: Exp Setup}(a).

The pyrometry has to be calibrated to derive the instrument factor as needed for quantitative temperature calculation \cite{Levendis.1992.Rev.Sci.Instrum}.~The calibration was performed with a NIST-certified tungsten strip lamp operated between 1346–2915K and known spectral emissivity of tungsten.~The entire optical settings were kept identical for calibration and combustion experiments, in which the exposure times are optimized for different AIR conditions to avoid signal saturation.~The uncertainty caused by the calibration was about 25\,K at a bulk temperature of 3100\,K.~Detailed calibration and temperature quantification processes are described elsewhere \cite{Ning.2024}.

% DBI Nano
As discussed earlier, the formation of nanoparticle clouds has been a crucial issue for iron combustion as undesired mass loss that impairs the recyclability in a reduction-oxidation scheme.~As a primary objective, this study aims to experimentally resolve the formation process of NP clouds.~For this purpose, a second DBI system (DBI Nano) was employed, as shown in Fig.~\ref{Fig: Exp Setup}(b).~Here, a high-speed CMOS camera (Phantom TMX 7510) was used, while all other optical components (LED, microscope, and filter) were identical to the DBI Sizing system.~To this end, a pixel resolution was aligned to a similar value of 5.2\,\textmu m.~The height of the FOV was adjusted for different AIR conditions, ensuring to temporally capture the NP clouds initiation process.~Similar to the DBI sizing experiments, the captured nanoparticle intensity is normalized to the homogeneous light background and then inverted; thus, $I_\textrm{nano}$ represents the strength of light attenuated by the NP clouds.

The three high-speed imaging systems were temporally synchronized at a 10\,kHz repetition rate using a programmable timing unit (PTU, LaVision).~The correlation of different physical parameters measured by the four cameras for the same individual particle was achieved through an in-house algorithm, which was extended based on the method described in \cite{Ning.2022.Combust.Flame}.~As a result, each particle's trajectory includes time-resolved data on particle diameter ($d_\textrm{prt}$), surface temperature ($T_\textrm{prt}$), and NP cloud intensity ($I_\textrm{nano}$).~The reference time $t_0$ corresponds to the moment when the particle passes through the main reaction zone of the flat flame, marking the onset of heating \cite{Li.2021.Fuel}.~In total, approximately 770 particles were measured under three different AIR conditions, allowing for double conditioning on initial particle size and temperature in the statistical analysis.

\subsection{Particles size divisions}

In this study, we utilized near-spherical iron particles provided by Eckart GmbH (99.8\% purity), with a sphericity (SPHT3 = 0.9) characterized using a CamSizer X2.~Prior to the experiments, the particles were sieved to a size range of 32–65\,\textmu m, and their statistical particle size distribution (PSD) was determined \textit{ex-situ} using the CamSizer X2.~Particle size was also evaluated from shadow images captured by the \textit{in-situ} DBI sizing technique, following the light attenuation approach first introduced in \cite{Ning.2022.Proc.Combust.Inst.} and further optimized in our previous work \cite{Ning.2023}.~This method calculates the circle-equivalent diameter ($d_\textrm{prt}$) by correlating the integrated light attenuation with the particle surface area, offering lower uncertainty and higher precision compared to gradient-based methods \cite{Li.2021.Fuel, Li.2022.Combust.Flameb, Li.2024}, as confirmed by verification against a high-precision dot pattern.

\begin{figure}[h!]
\centering
\includegraphics[width=75mm]{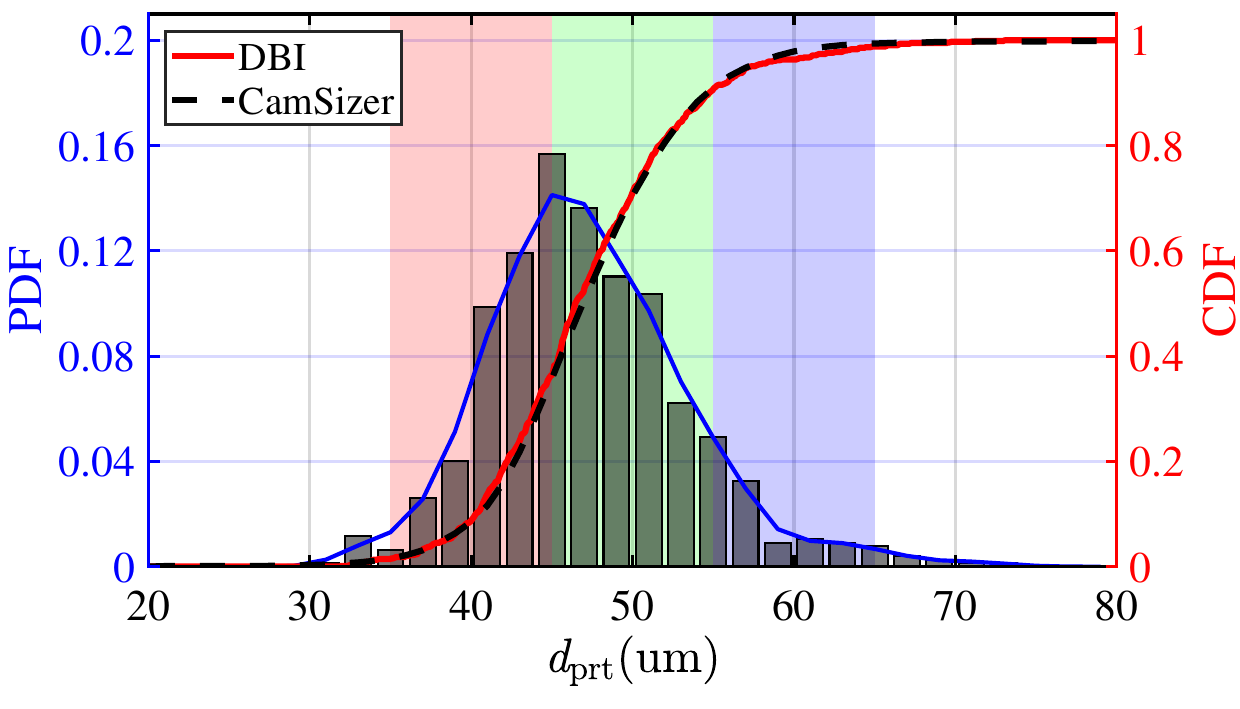}
   \caption{Probability density function (PDF, left axis) and cumulative density function (CDF, right axis) from \textit{in-situ} DBI sizing (red and blue solid lines) and CamSizer (red dashed line) measurements.}
\label{Fig: dp_PDF}
\end{figure}

Subsequently, the probability density function (PDF) of \textit{in-situ} $d_\textrm{prt}$ is evaluated, as shown in Fig.\,\ref{Fig: dp_PDF} (left y-axis).~Note that $d_\textrm{prt}$ represents the initial particle size during convective heating, which will be discussed further in Section \ref{subsec:parametertrajectories}.~Most particles fall within the diameter range of 30–70\,\textmu m, with the peak probability around 45\,\textmu m.~The cumulative probability density function (CDF), indicated by the right y-axis, shows excellent agreement between the \textit{in-situ} and \textit{ex-situ} size measurements.~Additionally, Table~\ref{Tab: Particle size divisions} compares three percentile diameters -- $d_\textrm{10}$, $d_\textrm{50}$, and $d_\textrm{90}$ -- with deviations between the two measurement methods all below 1\%, thereby verifying the high precision of the \textit{in-situ} DBI sizing method in determining particle diameter.

To perform statistical analysis conditioned on particle size, all measured particles are divided into three subgroups based on the \textit{in-situ} measured $d_\textrm{prt}$.~The diameter range of three subgroups, named as A, B, C, are centered at 40, 50, and 60 \textmu m, respectively, which are also color-coded in Fig.\ref{Fig: dp_PDF}.~Table \ref{Tab: Particle size divisions} provides more details on the sub-group range, particle numbers $N_\textrm{prt}$ and mean particle diameters.~In addition to the total number, the exact number of particles measured in each AIR conditions are also given by triples of numbers in brackets.~As a consequence of the size classification, less particles are present in the group C, which might lead to larger statistical uncertainties.~However, the numbers are comparable between three AIR conditions, and thus are deemed sufficient for comparison purposes. 

% Fig: PSD

\begin{table}[]
\centering
\small
\begin{tabular}{cccc}
\hline
Techniques  & $d_{10}$ (\textmu m)      & $d_{50}$ (\textmu m)     & $d_{90}$ (\textmu m)    \\ \hline
CamSizer   & 40.54                  & 46.92                 & 54.59     \\
DBI Sizing & 40.32                  & 46.49                 & 54.80     \\
Deviations  & 0.54\%                 & -0.92\%               & 0.38      \\ \hline
\hline
Group      & A                      & B                     & C         \\ 
\hline
Range (\textmu m)        & {[}35-45) & {[}45-55) & {[}55-65) \\ \hline
$N_\textrm{prt}$        &  \makecell{280 \\ (52/133/95)} & \makecell{418 \\ (173/127/154)} & \makecell{73 \\ (26/20/27)} \\
\hline
Mean $d_\textrm{prt}$   & 41.4   & 49.0 & 59.6\\
\hline
\end{tabular}
\caption{Comparison of percentiles based diameters $d_{10}$, $d_{50}$, and $d_{90}$ between the \textit{in-situ} sizing and the prior CamSizer X2 measurements and the division of three particle size groups.}
\label{Tab: Particle size divisions}
\end{table}

\section{Simulation tools}
\label{sec:simulation tools}

\subsection{CFD simulations}
\label{subsec:CFD simulation}

% \textcolor{Red}{This part will be contributed by Didi. Here, we need validation; refer to another paper or add a figure showing validation against Daoguan's previous temperature data.}

Single particle simulations with a resolved boundary layer were conducted to gain a more detailed insight of the processes in the boundary layer.~A detailed description of the computational setup can be found in~\cite{nguyen2024}.~In the following, the setup is briefly summarized.

\begin{figure}[h!]
    \centering
    \includegraphics[width=75mm]{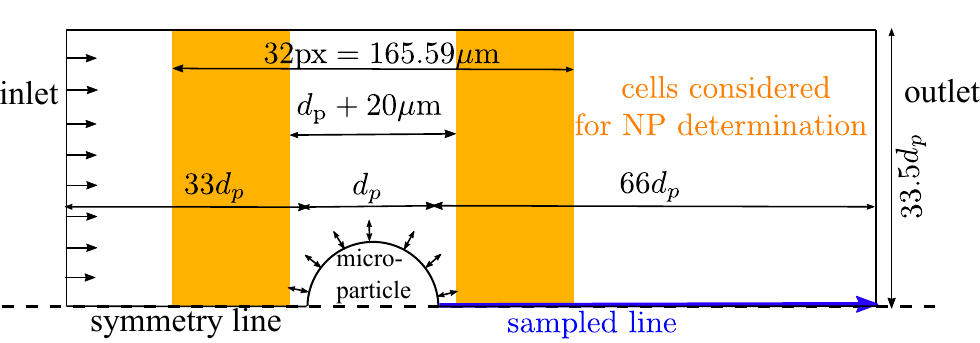}
    \caption{Sketch of the computational domain.~The orange frame depicts the region for integration of NP volume fraction.~The blue line depicts the line, at which the NP formation rate was sampled.}
    \label{fig:sim_domain}
\end{figure}

A 2D axisymmetric domain was used~(Figure~\ref{fig:sim_domain}), where the boundary layer is fully resolved and the particle is represented by a lumped model.~The particle model is a combination of the model by Mi~et~al.~\cite{Mi.2022} for the solid phase and the model by Mich~et~al.~\cite{Mich.2024} for the liquid phase.~A two-phase Eulerian-Eulerian approach was considered to capture the gas phase and the NP phase.~The NP phase transport equation is:

\begin{equation}
\frac{\partial\left(\alpha_\mathrm{c} \right)}{\partial t}+\frac{\partial\left(\alpha_\mathrm{c} u_i \right)}{\partial x_i} + \frac{\partial\left(\alpha_\mathrm{c} V_{\mathrm{c},i}^T \right)}{\partial x_i} = \frac{\alpha_\mathrm{g}\dot{\omega}_\mathrm{c}}{\rho_\mathrm{c}},
\end{equation}

where the subscript c denotes the condensed NP phase and the subscript g denotes the gas phase.~$\alpha$ is the volume fraction, $u_i$ is the convection velocity vector, $\dot{\omega}_\mathrm{c}$ is the NP production term, and $\rho_\mathrm{c}$ is the NP density.~$V_{\mathrm{c},i}^T$ is the thermophoretic velocity of the NPs, given by~\cite{friendlander2000}:

\begin{equation}
    V_{\mathrm{c},i}^T= -0.55 \frac{\nu_\mathrm{g}}{T} \frac{\partial T}{\partial x_i},
\end{equation}

with $\nu_\mathrm{g}$ as the kinematic viscosity. The Stefan flow, which can influence the detection of the NPs (discussed in Section~\ref{subsec:NP growth}), was applied at the particle surface:

\begin{equation}
u_\mathrm{Stefan}=n_i \frac{(\dot{m}_\mathrm{ox}+\dot{m}_\mathrm{evap})}{A_\mathrm{p}\rho_\mathrm{s}}.
\label{eq:stefan}
\end{equation}

where $n_i$ is the surface normal vector and $A_\mathrm{p}$ is the particle surface area.~The Stefan flow velocity is a function of the sum of oxygen consumption mass flux $\dot{m}_\mathrm{ox}$ and evaporation mass flux $\dot{m}_\mathrm{evap}$.~Note that the oxygen consumption mass flux is always going towards the particle and the evaporation mass flux is always going outwards.~Thus, the direction of the Stefan flow is determined by the more dominant process.

The precursors for the NP formation originate from the evaporation of the liquid microparticle.~To model evaporation, the partial pressure of Fe(g), obtained from chemical equilibrium calculations~(Figure~\ref{Fig: VaporPressure}), was prescribed at the microparticle surface.~The evaporation of other species were not considered, since the partial pressure of Fe(g) is several order of magnitudes higher than the partial pressure of other species.~The NP production $\dot{\omega}_\mathrm{c}$ was modeled as the condensation of gaseous iron species upon supersaturation, which aligns with NP formation models for alumina~\cite{finke2024,panda1995} and silica~\cite{janbazi2019}.

The determination of NPs was performed in an equivalent manner to the experimental measurements.~The NP volume fraction was integrated and normalized in a defined region shown in Figure~\ref{fig:sim_domain}, giving $\Sigma V_\textrm{nano, norm}$.~In streamwise direction, the frame length was chosen equal to the size of the field of view.~In line-of-sight direction, the entire height of the domain was used for integration.~A gap at the microparticle was left out, which accounted for the NPs that were close to or above the microparticle, that could not be detected experimentally.~$t_\mathrm{nano}$ was determined at the location of maximum curvature of $\Sigma V_\textrm{nano, norm}$, see Fig.\,S2 in the supplementary material.~To investigate the NP formation, the source term $\dot{\omega}_\mathrm{c}$ was sampled along a line from the microparticle surface in streamline direction (blue line in Figure~\ref{fig:sim_domain}).

\subsection{Molecular dynamic simulations}

To gain a basic understanding of the nanoparticle formation mechanism, we performed molecular dynamic simulations using reactive force fields (ReaxFF).~ReaxFF-MD is an empirical molecular dynamics potential formalism rooted in bond-order principles, meticulously parameterized against either quantum mechanics (QM) calculations or experimental measurements.~ReaxFF dissects the total energy of a system, denoted as $E_{system}$, into various energy components, as shown in Eq.~\ref{Eq:EnergyEquation} \cite{vanDuin.2001}:

\begin{equation}
    \begin{aligned}
	E_{system}  & = E_{bond} + E_{val} + E_{tors} \\
                & + E_{over} + E_{vdWaals} + E_{Coulomb} + E_{specific}.
    \end{aligned}   
        \label{Eq:EnergyEquation}
\end{equation}	

Here, $E_{bond}$, $E_{val}$, and $E_{tors}$ are the bond energy, valence angle energy, and torsion angle energy, respectively.~$E_{over}$ accounts for the corrective energy penalty arsing from over-coordination.~$E_{vdWaals}$ and $E_{Coulomb}$ denote the bond-order independent van der Waals and Coulomb energy contributions, respectively.~$E_{specific}$ represents system-specific energy contributions including but not limited to Ione-pair, conjugation and hydrogen-bonding corrections.~The seamless transition in bond order during bond formation and dissociation endows ReaxFF with the capability to accurately simulate chemical reactions.~Moreover, the electro-negativity equalization method (EEM) is employed by ReaxFF to calculate the atomic charges dynamically.~For a more comprehensive introduction about ReaxF, readers are referred to previous literature \cite{Senftle.2016,Uene.2019}.

In the present work, the employed Fe/O force field parameters are from the work of Shin et al.~\cite{Shin.2015}, which were trained with the thermodynamics of iron oxides and the energetics of iron redox interactions.~This force field, according to Thijs et al.~\cite{Thijs.2023}, provides a good prediction of liquid iron and iron-oxide mixing enthalpies, as well as coordination numbers in the mixing phase.~Hence, we anticipate that this force field offers distinct advantages over other force fields in probing iron-oxide composition. 

\begin{figure*}[h!]
\centering
\includegraphics[width=100mm]{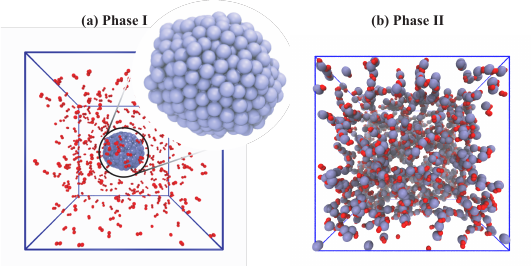}
   \caption{(a) Phase I: The initial configuration of the ReaxFF-MD simulation box consisting of a 500-atom Fe(s) particle and 400 \ce{O2} molecules.~(b) Phase II: The initial configuration of the ReaxFF-MD simulation box consisting of 1000 \ce{FeO2}(g) molecules.~Atom colors: Fe (ice blue), and O (red).}
\label{Fig: MDConfig}
\end{figure*}

Another challenge for our modeling is the timescale.~According to previously reported work \cite{Thijs.2023, Rumminger.1999, Nanjaiah.2021}, iron combustion involves the transition from Fe(l) to Fe(g) and FeO(g), which contribute to the formation of \ce{Fe2O3}(s).~Admittedly, this entire process is beyond the size and time-scales accessible to ReaxFF, which is typically limited to nanoseconds and nanometers.~Therefore, capturing the entire process of many precursors evaporating from the iron micro particle and gradually forming an iron-oxide nanoparticle is challenging.~Thus, we delineated the process of iron-oxide nanoparticle production into two phases I and II, as depicted by the configurations in Fig.\,\ref{Fig: MDConfig}.

% Fig: MD Configuration

The first phase (Phase I) involves the formation of \ce{FeO2}(g) precursors.~To achieve this, we prepared an iron powder model by annealing a BCC-Fe spherical particle with a radius of 12\,$\r{A}$.~Subsequently, the Fe particle was placed into an $80\r{A}\times 80 \r{A} \times 80 \r{A}$ simulation box with periodic boundary conditions, alongside 400 oxygen molecules (\ce{O2}), as shown in Fig.\,\ref{Fig: MDConfig}(a).~Following an energy minimization for the system, MD simulations were conducted with an initial temperature of 1500\,K for 200\,ps with the microcanonical ensemble, where the number of atoms (N), the simulation box volume (V), and the energy (E) are conserved.~Atomic motions were tracked using the velocity Verlet integration algorithm with a time step of 0.1\,fs.~Atomic charges were calculated at every step.~It is worth noting that the Fe particle is significantly smaller than the micro-powders in experiments.~Nevertheless, the ReaxFF-MD simulation aims to capture the atomic level reactions for producing iron-oxide nanoparticles, which typically entail only a limited number of atoms.~Besides, based on the ReaxFF-MD simulations of other but similar systems \cite{Gao.2022}, the size or shape of iron powders has minimal influence on the chemical reaction pathways.~Thus, the Fe particle consisting of 500 atoms is deemed sufficient for elucidating chemical reaction kinetics at the atomic scale, gaining a fundamental understanding of the precursor formation pathways.

In the second phase (Phase II), we dispersed 1000 \ce{FeO2} into an $80\r{A}\times 80 \r{A} \times 80 \r{A}$ simulation box using periodic boundary conditions, as shown in \ref{Fig: MDConfig}(a).~The energy-minimized configuration underwent MD simulations at 1000\,K and 2000\,K for 250\,ps using the canonical ensemble, where the number of atoms (N), the simulation box volume (V), and the temperature (T) were conserved.~The time step for MD-NVT simulation was set to 0.25\,fs, with atomic charges updated at each step.~Moreover, the Nose-Hoover thermostat is employed to regulate the temperature throughout the MD simulations.

\section{Results and Discussions}
\label{sec:R&D}

\subsection{Trajectories of particle size, nanoparticle clouds, and temperature}
\label{subsec:parametertrajectories}

\begin{figure}[h!]
\centering
\includegraphics[width=70mm]{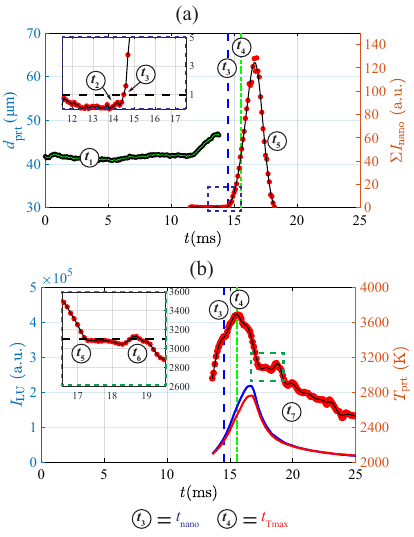}
   \caption{Tracking the combustion of a single iron particle using multi-parameter diagnostics.~(a) Particle diameter $d_\textrm{prt}$ and nano-particle signal integration $\Sigma I_\textrm{nano}$.~(b) Luminosity intensities $I_\textrm{LU}$ of two pyrometry channels (blue and red lines) and derived microparticle surface temperature.}
\label{Fig: SingleShot_Trajact}
\end{figure}

\begin{figure*}[h!]
\centering
\includegraphics[width=140mm]{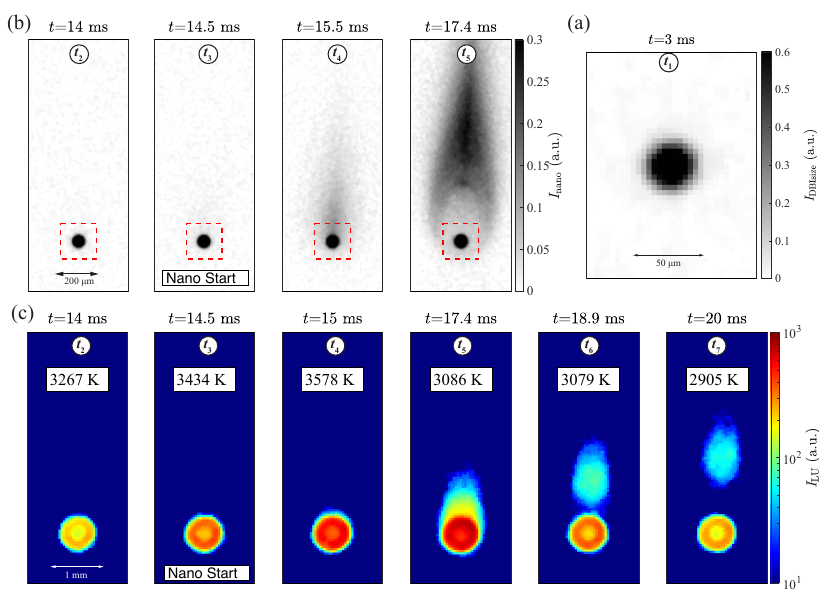}
   \caption{Images at selected time steps.~(a) Typical particle shadow image at $t=3$\,ms from the DBI sizing measurement.~(b) Particle and NP shadow images from the DBI Nano measurement.~(c) Luminosity images of the 850\,nm channel with evaluated temperature from the 2C-pyrometry measurement.}
\label{Fig: SingleShot_Image}
\end{figure*}

One key question driving the experimental efforts is how multi-parameter diagnostics can deepen current understanding of the oxidation process of a single iron particle, or what insights can be gained about the underlying physics by combining time-resolved scalar measurements.~To explore the potential—and recognize the limitations—offered by advanced diagnostic techniques, this section presents examples of individual particle trajectories along with the associated time-resolved scalar data.~As a demonstration, we examine the case of a 40\,\textmu m particle (size group A) burning in an AIR40 atmosphere (40vol\% \ce{O2}).~The temporal evolution of different parameters are depicted in Fig.~\ref{Fig: SingleShot_Trajact}, with corresponding images at selected time steps shown in Fig.~\ref{Fig: SingleShot_Image}.

% Describe the multi-para trajectory of a particle in AIR40. Combine Figure 5 whenever it is necesary. Add two normal particle from AIR20, and AIR30 in SM (tbd)

% Point 1. Particle diameter: How to define and how it changes
During the initial stage, the particle diameter $d_\textrm{prt}$, shown in Fig.~\ref{Fig: SingleShot_Trajact}(a) (black dots, refer to the left y-axis), remains nearly constant.~Since the particles are almost spherical, as depicted in Fig.~\ref{Fig: SingleShot_Image}(a) at $t = 3$\,ms ($t_1$), rotational motion of the particles has minimal impact on determining the projected particle area.~The particle size is influenced by mass gain, which progresses slowly in the kinetic-limited regime, as the particle temperature increase is primarily caused by convective heating.~Given that the particle temperature is expected to be below its melting point, this stage is referred to as the solid-phase oxidation stage \cite{Ning.2023}, extending approximately from 0 to 10\,ms for this particle.~Subsequently, an increase in $d_\textrm{prt}$ is observed, attributed to the density decrease (solid to liquid phase transition) and mass gain through oxidation \cite{Ning.2022.Proc.Combust.Inst.}.~To reduce uncertainty due to the beam steering effect of the flat flame and the considerable size growth during the diffusion-limited combustion, the particle diameter measured between 3 and 6\,mm HAB is averaged to represent the initial particle diameter; this procedure is applied to each individual particle.

% Fig: A large figure shows two curves and Nano and T images

% Point 2. Nanoparticle cloud: how to detect and define initiation time

The NP intensity integration $\Sigma I_\textrm{nano}$ is obtained by integrating $I_\textrm{nano}$ within the dashed box (see Fig. \ref{Fig: SingleShot_Image}(b)) but excluding the contribution of the microparticle.~By capturing its dynamic process, it becomes feasible to determine the apparent initiation time of NP formation.~Specifically, the scattered $\Sigma I_\textrm{nano}$ data is first fitted using a spline function with constrained freedom \cite{Li.2022.Proc.Combust.Inst.b}.~A fixed threshold is then applied to intersect the fitted curve, which allows for determining the NP initiation time, $t_\textrm{nano}$.~In the zoomed-in view of Fig.~\ref{Fig: SingleShot_Trajact}(a), $\Sigma I_\textrm{nano}$ remains near zero before NP initiation (e.g., at $t_2$), then rapidly increases at $t_3$ ($t_3 = t_\textrm{nano}$).~While an alternative method could define $t_\textrm{nano}$ using the minimum of the second derivative of $\Sigma I_\textrm{nano}$, the difference is found to be minor.~The sharp gradient in time-resolved $\Sigma I_\textrm{nano}$ effectively reduces uncertainty in determining $t_\textrm{nano}$.

% Point 3. Temperature: evolution and plateau
The initial time of a detectable particle surface temperature $T_\textrm{prt}$ or the minimum $T_\textrm{prt}$ depends on the camera exposure time, which varies in the three AIR conditions.~As mentioned previously, the exposure time is adjusted to capture the maximum particle temperature by avoiding saturating the camera sensor.~In Fig.~\ref{Fig: SingleShot_Trajact}(b), the detected luminosity intensities, $I_\textrm{LU}$, are shown for the 950\,nm channel (red) and the 850\,nm channel (blue) alongside the evaluated particle temperature.~A rapid increase in $I_\textrm{LU}$ is observed, reaching its peak near $t_4$ ($t_4=t_\textrm{Tmax}$), driven by the heat released from exothermic oxidation reactions.~Although convection and radiation heat losses are significant \cite{Ning.2024}, they are not yet dominant at this stage.~\textit{Ex-situ} characterization of the oxidation stage suggests that \ce{FeO}(l) has predominantly formed by the time $T_\textrm{prt}$ is reached, accounting for approximately 75\% of the total mass \cite{Sperling.2025}.~Following this peak, $T_\textrm{prt}$ decreases as further oxidation from \ce{FeO}(l) to \ce{Fe3O4}(l) proceeds more slowly, and heat loss becomes increasingly important during the reactive cooling process \cite{Mich.2024}.~This phase, spanning from melting to re-solidification, is commonly refereed as the liquid-phase oxidation stage, where oxygen diffusion to the particle surface is one of the rate-limiting processes.

% Fig: A large figure shows two curves and Nano and T images

% Potential influnce of nanoparticles
The NP clouds introduce potential bias in the particle temperature evaluation from the time starting at NP initiation ($t_3$) until detachment from the parent particle ($t_6$), as visualized in Fig.\ref{Fig: SingleShot_Image}(c).~Specifically, there are two major limiting factors in the experiment.~First, the 2C-pyrometry is a line-of-sight measurement, and NPs along the observation direction and those attached to the parent particle surface are impossible to distinguish.~This could potentially overestimate the peak temperature due to the different emissivity of nano and micro iron oxides \cite{JosephKalman.2015}.~Second, as a defocusing of both pyrometry cameras was necessary to enhance the SNR, the luminosity of NPs can only be separated when they are significantly distant from the parent microparticle.~As NP clouds grow and move downstream (due to higher velocity), the luminosity image becomes elongated, as seen at $t_5$ in Fig.\ref{Fig: SingleShot_Image}(c).~In such cases, the luminosity ratio over-represents the real surface temperature.~Depending on the volume fraction of NPs, this bias often leads to a slowly decreasing temperature \cite{Ning.2024} or a temperature plateau, as shown between $t_5$ and $t_6$ in the zoomed-in view of Fig.\ref{Fig: SingleShot_Trajact}(b).

We believe that this experimental bias is unavoidable due to the line-of-sight nature of the measurements, and caution must be taken when comparing experimentally determined particle temperatures with simulations.~However, we contend that this uncertainty is only relevant for conditions where NP volume fractions are high, namely AIR40 in this study, while it is much less pronounced in AIR30 and negligible in AIR20.~Nevertheless, the following discussion will focus on the temperature before and at NP initiation time, rather than emphasizing the peak temperature.

\subsection{Size and oxygen-dependent particle temperature}
\label{subsec:particletemperature}

\begin{figure}[h!]
\centering
\includegraphics[width=75mm]{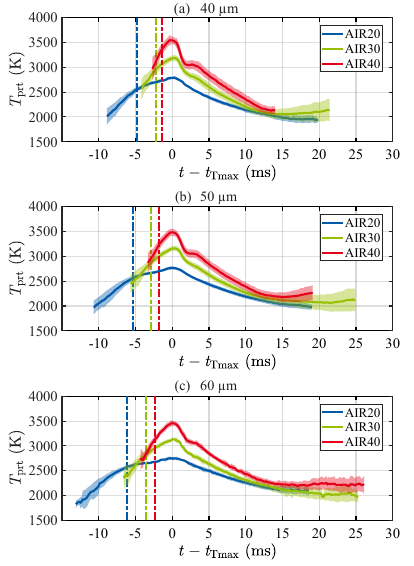}
   \caption{Temporal evolution of the particle temperature $T_\textrm{prt}$ for three particle size groups with mean $d_\textrm{prt}$ at about (a) 40 \textmu m, (b) 50\textmu m, and (c) 60 \textmu m, which are conditioned at the maximum temperature time $t_\textrm{Tmax}$.~Vertical dashed lines indicate the nanoparticle initiation time $t_\textrm{nano}$
   }
\label{Fig: T-Profiles}
\end{figure}

In this section, we analyze the temporal history of the particle temperature of the three subgroups centered at about 40, 50, and 60\,\textmu m.~These profiles are further divided into three AIR conditions and aligned to the time of peak temperature $t_\textrm{Tmax}$.~Applying this double conditioning, Figure~\ref{Fig: T-Profiles} depicts the mean (solid lines) and $\pm$ one standard deviation (shaded error bars) of the particle temperature.~Vertical dashed lines indicate the mean time, $t_\textrm{nano}$, of NP initiation, which will be elaborated in Section~\ref{subsec:NP initiation}.

A weak correlation between particle size and particle temperature is observed, which is due to the larger radiative and evaporative heat loss associated with a larger diameter and has been experimentally and theoretically analyzed in our previous work \cite{Ning.2024}.~In contrast, particle temperature deviates significantly among the three AIR conditions and increases at higher \ce{O2} concentrations.~This can be well explained as the particle burns in a diffusion-limited regime \cite{Ning.2023}, in which the reaction rate is controlled by the oxygen transport to the particle surface, driven by diffusion and Stefan flow.~Thus, higher ambient \ce{O2} mole fraction leads to faster oxidation reactions and heat release, elevating the surface temperature.

% Discuses how the measured temperature develops with existing nanoparticles, refer to the AIR40 single shot
Nonphysical bias caused by NP can be also observed in these conditioned particle temperature profiles.~The peak temperature in AIR40 reaches about 3500\,K, which is even higher than the vaporization-dissociation temperature of FeO \cite{Yetter.2009.Proc.Combust.Inst.}.~We consider that it is nonphysical as the dissociation reaction limits the highest reachable temperature, and the artifacts likely stem from the different emissivity of NPs.~In addition, an nonphysical temperature plateau in AIR40 is evident at about 2-3\,ms after the peak temperature, while significantly phased out in AIR30 and eliminated in AIR20.~Nevertheless, double-conditioned temperature profiles reveal the importance of oxygen availability for the diffusion-limited oxidation, and quantitative assessment is still feasible whenever the NP volume is negligible, which is valid for the entire AIR20 and AIR30, and before $t_\textrm{nano}$ in AIR40. 

\subsection{Nanoparticle initiation}
\label{subsec:NP initiation}

\begin{figure}[h!]
\centering
\includegraphics[width=75mm]{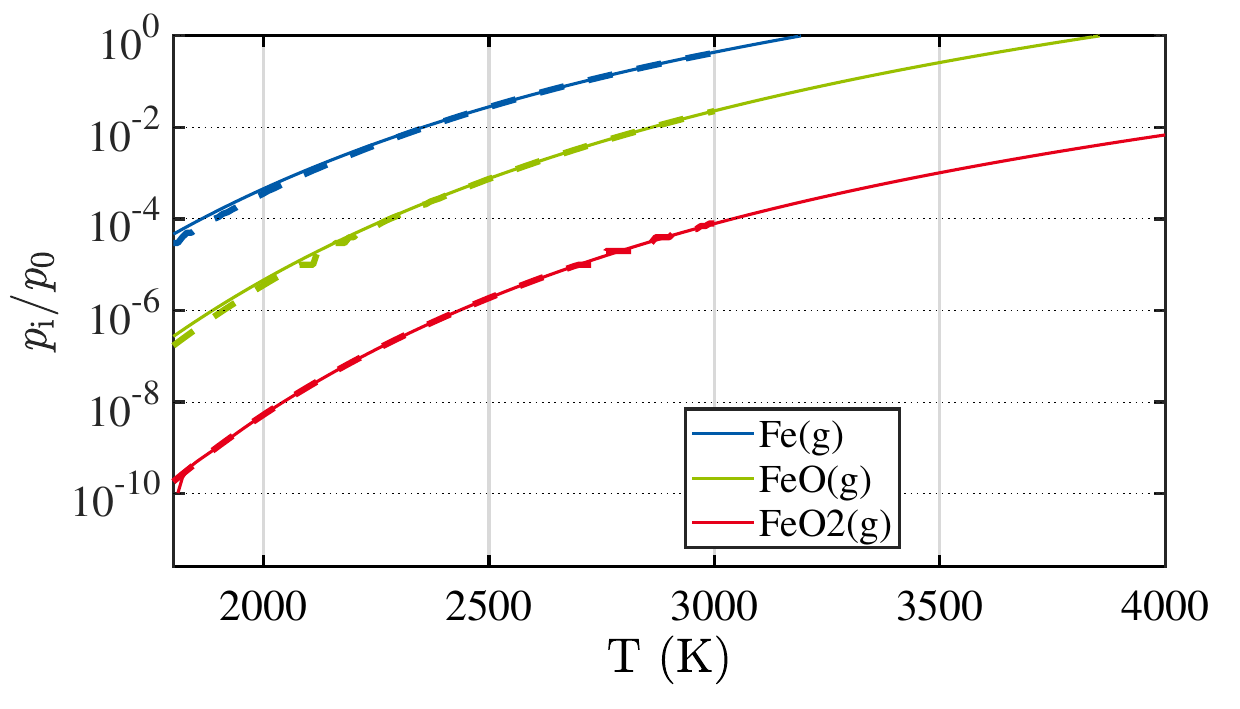}
   \caption{Partial pressure of gas-phase Fe(g), FeO(g), and \ce{FeO2}(g) over temperature in chemical equilibrium with an atomic Fe:O ratio set to 8:2.}
\label{Fig: VaporPressure}
\end{figure}

% chemical equilibrium calculations of vapor pressure
The formation of nanoparticle (NP) clouds is strongly influenced by precursors, which are formed in the gas phase or directly released from the condensed phase into the gas phase \cite{Thijs.2023}.~To understand this process, chemical equilibrium calculations were first performed to determine the partial pressures $p_i/p_0$ of gas-phase compositions Fe(g), FeO(g), and \ce{FeO2}(g), as shown in Fig.~\ref{Fig: VaporPressure}.~These calculations were conducted over a temperature range of 1600 to 3000\,K (indicated by dashed lines), with polynomial fitting applied for extrapolation beyond this range.~Among the species analyzed, Fe(g) exhibits the highest partial pressure, surpassing FeO(g) by approximately two orders of magnitude and \ce{FeO2}(g) by five orders of magnitude.~The Fe:O mole ratio was initially set to 8:2 as a boundary condition, but the results remained nearly consistent across varying Fe/O ratios.~The partial pressures of these precursors are notably temperature-dependent, increasing monotonically with rising temperature, but largely independent on the composition of the liquid.~Assuming that NPs form through the recombination and aggregation of gas-phase iron and iron oxide precursors, it can be hypothesized that NP formation is primarily governed by the evaporation and temperature of the condensed phase.

% Vapor pressure of Fe and FeO over Temperature

\begin{figure}[h!]
\centering
\includegraphics[width=75mm]{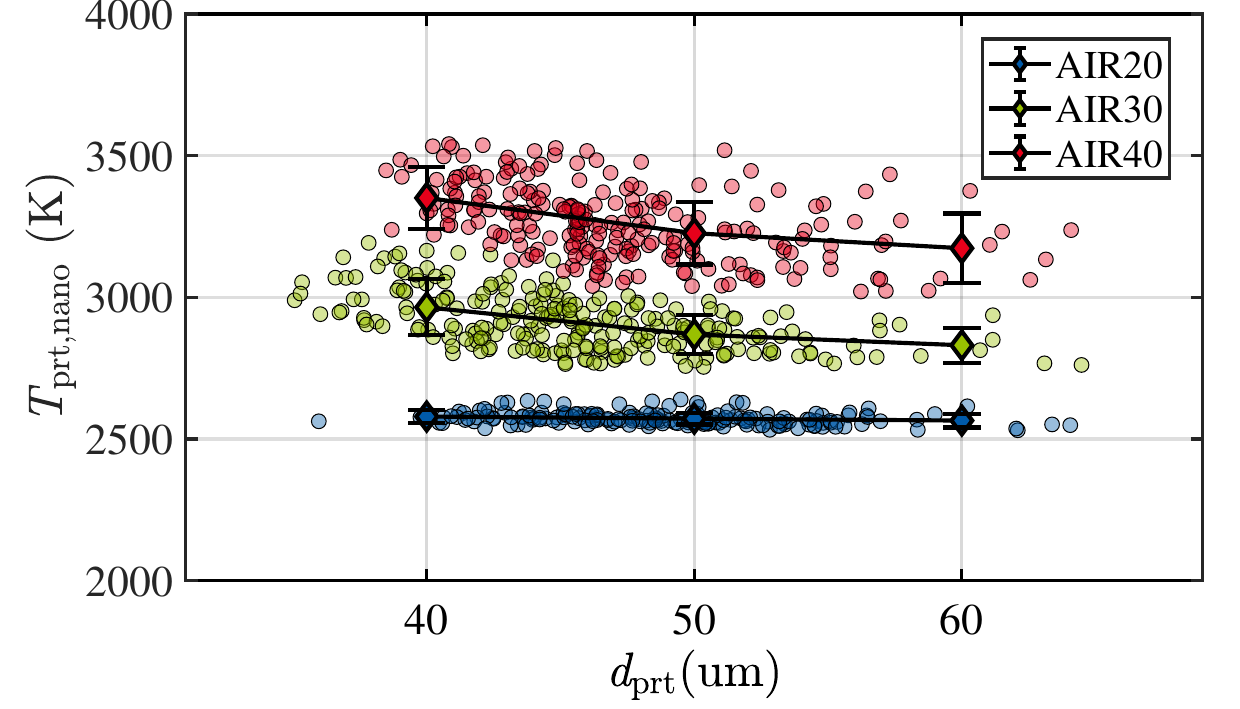}
   \caption{Scatter plot of particle temperature at NP initiation $T_\textrm{prt, nano}$ for all investigated conditions.~The mean and two standard deviations are also shown for three size groups.}
\label{Fig: Tnano}
\end{figure}

To experimentally test the hypothesis mentioned above, the microparticle surface temperature at the moment of nanoparticle initiation, $T_{\text{prt, nano}}$, is presented as a scatter plot in Fig.~\ref{Fig: Tnano}.~The mean values and two standard deviations are shown for three different size groups and three different AIR conditions.~Contrary to the initial hypothesis, the experimental data indicates that $T_{\text{prt, nano}}$ is not constant.~Instead, it displays a weak dependence on particle size and a strong dependence on the gas-phase \ce{O2} mole fraction.~However, the underlying mechanisms may differ.

As discussed earlier, the NP initiation time $t_{\text{nano}}$ is defined as the moment when NP clouds become detectable using the present DBI (nano) imaging method within the defined region of interest (ROI) around the parent particle, as illustrated in Fig.~\ref{Fig: SingleShot_Image}(b).~This detection criteria signifies that the NP volume has reached a minimum level given by an empirically determined threshold at $t_{\text{nano}}$.~Although this minimum concentration of NPs cannot be quantified with the current diagnostics, the comparison of $T_{\text{nano}}$ remains valid because the detection criteria have been consistently used during data processing.

In this context, the size dependency of the NP initiation temperature can be explained.~Larger particles have a greater surface area, which leads to the release of a larger quantity of iron and iron oxide precursors into the gas phase.~This increased availability of precursors promotes aggregation and clustering, enabling NP clouds to become detectable at an earlier stage or lower temperature.

% T_nano: critical temperature at nanoparticle initiation

However, understanding the dependency of $T_{\text{nano}}$ on $\ce{O2}$ is not that straightforward.~Firstly, $\ce{O2}$ plays a crucial role in the formation and agglomeration of precursors, a topic that will be further explored in Section \ref{subsec:precursor formation} using molecular dynamics simulations.~Changes in the gas-phase $\ce{O2}$ concentration will therefore influence the pathway of NP formation.~Secondly, a higher partial pressure of $\ce{O2}$ enhances the Stefan flow.~In the extreme case, convection forces dominated by Stefan flow might trap NPs on the surface or in the near-surface region, rendering them spatially unresolved.~This hypothesis is supplemented by previous SEM analyses of collected iron oxide products, which showed some NP clusters preserve on surfaces \cite{Sperling.2025}.~Considering the effect of the Stefan flow, a higher $\ce{O2}$ mole fraction could delay the release of NPs into the surrounding gas, leading to a higher $T_{\text{nano}}$.~Since measuring fluid motion near the particle is highly challenging, we employed spatially-resolved numerical simulations to further explore this aspect, as discussed in Section \ref{subsec:NP growth}.

% t_nano
\begin{figure}[h!]
\centering
\includegraphics[width=75mm]{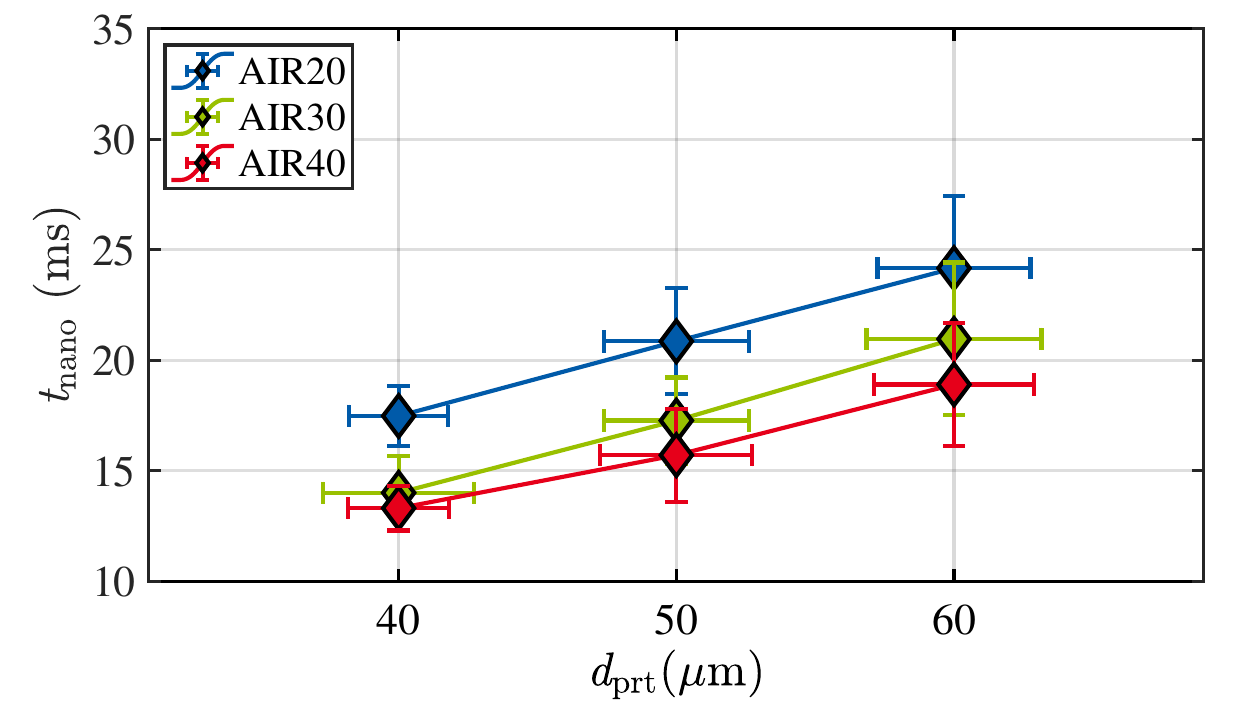}
   \caption{Nanoparticle initiation delay time $t_\textrm{nano}$ in dependency of \ce{O2} mole fraction and particle diameter $d_\textrm{prt}$.~Mean values are shown with error bars indicating two standard deviations for both $t_{\text{nano}}$ and $d_{\text{prt}}$.}
\label{Fig: tnano}
\end{figure}

Another key quantitative proxy of nanoparticle formation is the initiation time, $t_{\text{nano}}$.~Figure \ref{Fig: tnano} presents the mean NP initiation time for all conditions investigated, with error bars indicating two standard deviations for both $t_{\text{nano}}$ and $d_{\text{prt}}$.

Firstly, $t_{\text{nano}}$ in Fig.~\ref{Fig: tnano} exhibits a much stronger size dependency compared to $T_\textrm{prt, nano}$ shown in Fig.~\ref{Fig: Tnano}.~This is because larger particles experience a slower temperature increase, as previously shown in Fig.~\ref{Fig: T-Profiles}.~During the initial heating stage, the particle remains in a kinetically limited regime with minimal heat release from oxidation, so convective heat transfer dominates, making the particle heating rate almost inversely proportional to $d^2_\mathrm{prt}$.~After ignition, the particle burns in the diffusion-limited regime where the reaction rate is proportional to $d_\mathrm{prt}$ and thus the same 1/$d^2_\mathrm{prt}$ scaling is also approximately applied to the particle heating rate.~Consequently, a slower temperature rise during the entire oxidation process of larger particle requires a longer residence time to reach the NP initiation temperature.~Secondly, for similar particle size, Figure~\ref{Fig: tnano} suggests that NPs form earlier as the $\ce{O2}$ mole fraction increases.~The decrease in $t_{\text{nano}}$ can be attributed to the faster particle heating rate, as a higher $\ce{O2}$ mole fraction accelerates reaction progress and heat release in the diffusion-limited oxidation phase.~Thus, despite the higher $T_\textrm{prt, nano}$ at elevated $\ce{O2}$ mole fractions, $t_{\text{nano}}$ is still shorter.

However, the strong dependency of the nanoparticle initiation time $t_{\text{nano}}$ and the temperature $T_\textrm{prt, nano}$ on $X_{\ce{O2}}$ remains an open question that cannot be answered easily by experiments only.~To address this, a combined interpretation of experimental observations and numerical simulations was conducted.~The dynamic process of NP cloud growth will be discussed in the next section.

\subsection{Nanoparticle cloud growth}
\label{subsec:NP growth}

\begin{figure}[h!]
\centering
\includegraphics[width=75mm]{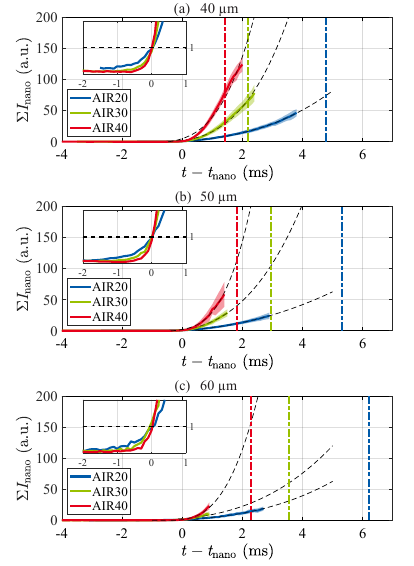}
   \caption{Temporal evolution of the nanoparticle intensity integration, $\Sigma I_{\text{nano}}$, for three particle size groups with mean $d_\textrm{prt}$ at about (a) 40 \textmu m, (b) 50\textmu m, and (c) 60 \textmu m, which are conditioned at the nanoparticle initiation time $t_\textrm{nano}$.~Vertical dashed lines indicate the additional time needed to reach the peak surface temperature.}
\label{Fig: Inano}
\end{figure}

In the first step, we analyze the temporal evolution of the measured NP clouds from the experiments.~Figure \ref{Fig: Inano} shows the integrated DBI intensity of the NP clouds, $\Sigma I_{\text{nano}}$, which is conditioned on the initiation time $t_{\text{nano}}$.~The solid lines represent the temporal mean of $\Sigma I_{\text{nano}}$, while the shaded areas indicate two standard deviations.~Additionally, vertical dashed lines mark the relative time of the peak surface temperature, $t_{\text{T,max}}$.~It is evident that reaching the peak temperature is delayed with increasing particle size and decreasing $\ce{O2}$ mole fraction.~This observation aligns with the earlier discussion regarding the overall slower rate of temperature rise.

Since NP clouds are only detectable for a brief period that is limited by the camera's field of view, $\Sigma I_{\text{nano}}$ is extrapolated for longer times using a polynomial fitting (dashed line) to better describe its trend and dynamics.~This fitting reveals that the development of NPs is more rapid at higher $\ce{O2}$ mole fractions.~Focusing on the initial stage in the zoom-in views, it becomes clear that a small portion of NPs have already formed before they are detected by the defined threshold (marked by the horizontal dashed line).~Assuming this threshold represents a specific total volume of NPs, the observations suggest that iron burning in lower $\ce{O2}$ environments requires more time to reach this specific total volume of NPs, whereas the NP clouds grow much quicker at higher $\ce{O2}$ concentrations.

\begin{figure}[h!]
\centering
\includegraphics[width=75mm]{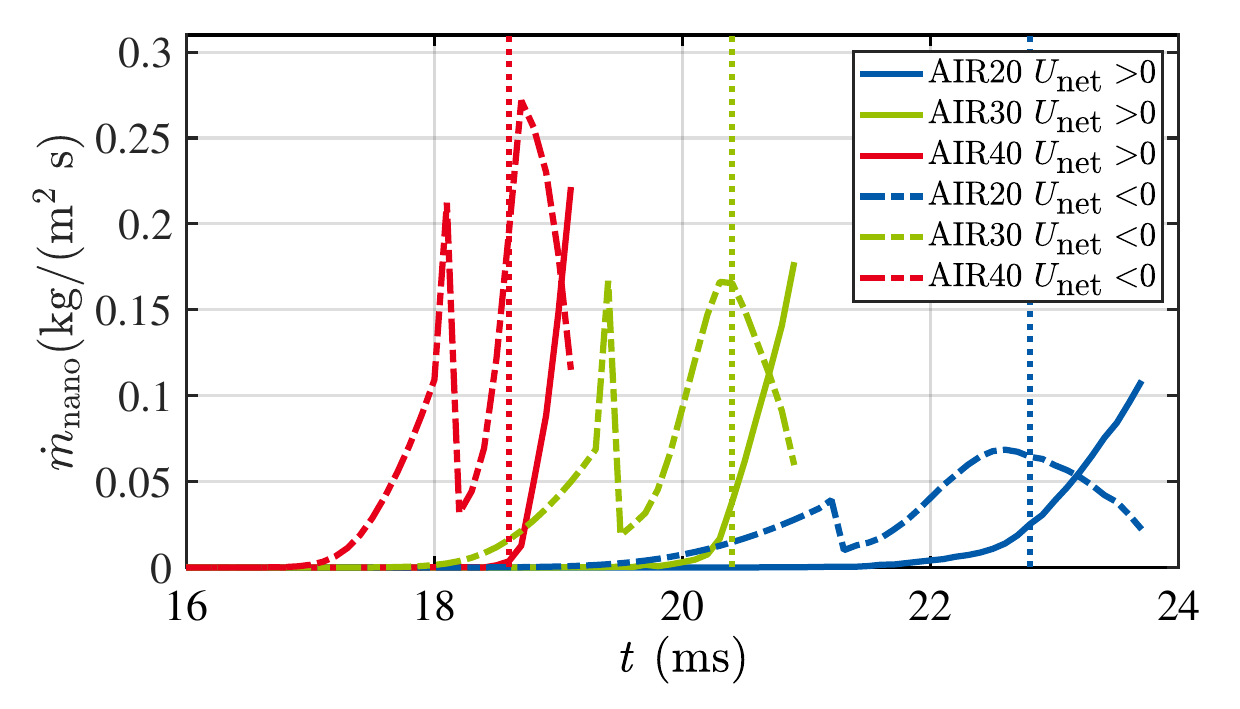}
   \caption{CFD simulation results: NP formation rate for a 50\,\textmu m particle conditioned on positive net velocity (solid lines) and negative net velocity (dashed lines).~NP initiation is shown by the vertical dotted lines.}
\label{Fig:Sim_NPProd_over_t_50um}
\end{figure}

To further understand this phenomenon, detailed numerical simulations are performed, as introduced previously in Section \ref{subsec:CFD simulation}.~Simulation results are validated against experimental data in terms of particle temperature history, see Fig.\,S1 in the supplementary material.~The NP formation rate sampled along a line~(Figure~\ref{fig:sim_domain}) is conditioned based on the net NP velocity, which is the sum of the convection velocity and thermophoresis velocity.~During diffusion-limited combustion, the hottest region is the microparticle surface.~Since thermophoresis transports NPs from hot to cold regions, the thermophoresis velocity is always positive (i.e. away from particle).~The convection velocity near the particle surface is mostly dominated by the Stefan flow, while further downstream it is mostly dominated by the particle slip velocity.~Conditioning the NP formation rate based on the net velocity gives an indication, whether the produced NPs are transported towards the particle ($U_\mathrm{net}<0$), or away from the particle ($U_\mathrm{net}>0$).~By performing this analysis, Figure \ref{Fig:Sim_NPProd_over_t_50um} shows the NP formation rate $\dot{m}_\textrm{nano}$ for a 50\,\textmu m particle over the residence time for the three investigated conditions.~Additionally, the numerical $t_\mathrm{nano}$ determined from $\Sigma V_\textrm{nano, norm}$ is depicted as vertical lines.~The temporal evolution of $\Sigma V_\textrm{nano, norm}$ is shown by Fig.\,S2 in supplementary material.

%\textcolor{red}{Sim: NP productivity for 50um}

%Considering that only the NPs transported away from, instead of towards, the particle surface lead to the growth of NP clouds, the numerical NP initiation time is calculated when $\dot{m}_\textrm{nano}$ at $U > 0$ reaches a specific threshold.~As indicated by the vertical dashed lines, NP initiation time is delayed in lower-oxygen conditions, which is consistent with the experimental observations in Fig.~\ref{Fig: tnano}.

For all AIR conditions it can be observed that $t_\mathrm{nano}$ (vertical dotted lines in Fig.\,\ref{Fig:Sim_NPProd_over_t_50um}) coincides roughly with the point of maximum curvature of the curve of $\dot{m}_\textrm{nano}$ at $U_\mathrm{net}>0$.~This indicates that only NPs that are transported away from the particle contribute to the cloud growth and can be detected by experimental measurements. Notably, NP initiation time is delayed in lower-oxygen conditions, which is consistent with the experimental observations in Fig.~\ref{Fig: tnano}.
%~Although the simulation slightly overestimates $t_\textrm{nano}$, this discrepancy is likely due to inaccuracies in the low-temperature oxidation mechanism.
Moreover, NP clouds evolve more rapidly (indicated by a steeper $\dot{m}_\textrm{nano}$ gradient) under higher-oxygen conditions, aligning with the experimental results for NP intensity integration $\Sigma I_\textrm{nano}$ shown in Fig.~\ref{Fig: Inano}.~In summary, the growth of NP clouds, as evaluated by $\dot{m}_\textrm{nano}$ at $U_\mathrm{net} > 0$ in the simulation, shows overall qualitative agreement with experimental findings.

Interestingly, NP formation rate at $U_\mathrm{net} < 0$ can be observed in the simulations before the experimentally apparent NP initiation, as depicted by the dashed curves in Fig.~\ref{Fig:Sim_NPProd_over_t_50um}.~This means that nanoparticles are already being produced prior to $t_\textrm{nano}$.~These particles, initially located near the microparticle surface, are transported towards the surface.~This is due to the fact that the negative Stefan flow velocity, caused by \ce{O2} consumption, outweighs the positive thermophoresis velocity.~As the temperature of the microparticle increases, the convection velocity resulting from the evaporation mass flux and the thermophoresis velocity increase accordingly, both directed away from the microparticle.~Therefore, eventually the region with positive net velocity grows, while the region with negative net velocity decreases.

The analysis of different AIR conditions reveals that the quantity of nanoparticles transported to the surface increases with higher \ce{O2} concentration due to the enhanced Stefan flow.~It clearly shows that NPs produced in the early phase of NP formation stay in the vicinity of the particle surface, making them undetectable by line-of-sight imaging methods used in the experiments.~In fact, NP deposite on microparticles collected from the combustion products was observed in many SEM analysis \cite{Ning.2022.Combust.Flameb, Sperling.2025}.~Such a process could lead to three consequences.~(1) The experimental $t_\textrm{nano}$ may be overestimated.~Comparing $\dot{m}_\textrm{nano}$ at $U_\mathrm{net}>0$ and $U_\mathrm{net}<0$, this overestimation ranges between 1 and 3 ms for a 50\,\textmu m particle.~(2) NPs attached to the surface could affect the radiation intensity due to their different emissivity.~We expect this effect to be minor since the quantity of NPs produced during the initiation time should still be small.~However, quantification of the emissivity change is not feasible with the current diagnostics.~(3) Most importantly, the delay in NP detection results in an overestimation of the experimental particle temperature at initiation, $T_\textrm{prt,nano}$.~Since $T_\textrm{prt}$ rises more rapidly in AIR40 than in AIR20, the measured $T_\textrm{prt,nano}$ is expected to be higher in high-oxygen environments.~The last point well explains the discrepancies in experimental $T_\textrm{prt,nano}$ observed in Fig.~\ref{Fig: Tnano}.

\begin{figure}[h!]
\centering
\includegraphics[width=75mm]{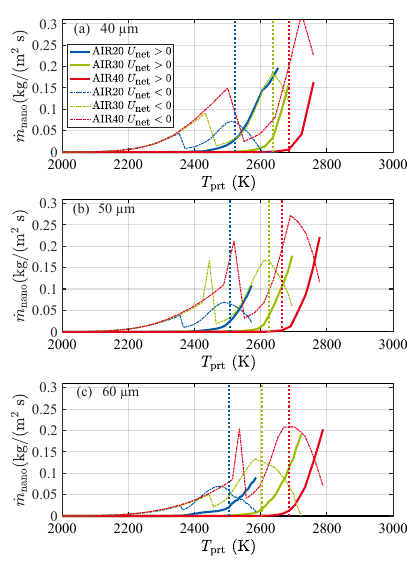}
   \caption{CFD simulation results: NP formation rate for (a) 40\,\textmu m, (b) 50\,\textmu m and (c) 60\,\textmu m particles conditioned on the net velocity $U_\textrm{net}$ (solid lines: positive, dashed lines: negative).}
\label{Fig:Sim_NPProd_over_T}
\end{figure}

% NP productivity over temperature
To gain a deeper understanding of the critical temperature for NP formation, Figure\,\ref{Fig:Sim_NPProd_over_T} presents NP formation rate, $\dot{m}_\textrm{nano}$, as a function of particle temperature, $T_\textrm{prt}$, for both positive (thick solid curves) and negative (thin dot-dash lines) net velocities.~It's important to emphasize that only positive net velocity leads to the detectable growth of NP particle clouds.~The formation rate at regions with negative net velocity start to rise at approximately 2200\,K, regardless of particle diameter or oxygen concentration.~As hypothesized in Section \ref{subsec:particletemperature} and confirmed by this simulation, it is not unexpected since the evaporation process is predominantly controlled by temperature, and NP formation occurs immediately as gas-phase Fe atoms are released.%~This will be further elaborated on in Section \ref{subsec:precursor formation}.

Figure\,\ref{Fig:Sim_NPProd_over_T} indicates that the simulated NP initiation temperature, $T_\textrm{prt,nano}$ slightly decreases with increasing particle diameter and significantly increases at higher oxygen concentrations.~Compared to the experimental results in Fig.\,\ref{Fig: Tnano}, the simulation is in very good quantitative agreement for 20 vol\% oxygen, where $T_\textrm{prt, nano}$ is approximately 2500\,K, while it underpredicts $T_\textrm{prt, nano}$ for 30 and 40vol\% oxygen.~This is probably due to multiple reasons including the emissivity of NPs attached to the surface and uncertainties of existing kinetic models in temperature predictions \cite{Nguyen.2024}.~Nevertheless, the trend that $T_\textrm{prt, nano}$ increases with oxygen concentration is evident in the simulations and experiments, which is well explained by the increased Stefan flow transporting NPs towards the microparticle surface.

\subsection{Precursor formation and agglomeration}
\label{subsec:precursor formation}

\begin{figure*}[h!]
\centering
\includegraphics[width=140mm]{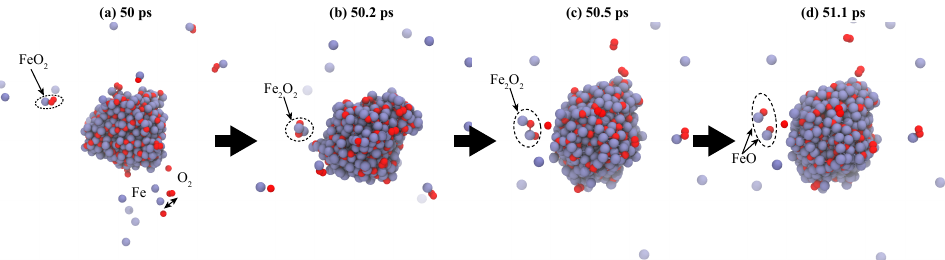}
   \caption{MD simulation (Phase I): precursor evolution from Fe(g) to \ce{FeO2}(g) and FeO(g).~Atom colors: Fe (ice blue) and O (red).}
\label{Fig: MDPrecursor}
\end{figure*}

The precursor formation pathways can be examined in Phase I using the MD simulation.~The temperature progression of the system is depicted in Fig.\,S3 of the supplementary material.~Due to the exothermic nature of Fe(s) oxidation, the temperature gradually rises to 4500 K.~Notably, around 3100\,K, the first dissociation of Fe(g) atoms from the Fe(l) particle is observed.~By the 100 ps mark, 16 Fe(g) atoms have been released from the initial Fe(l) particle.~The process of precursor formation is illustrated in Fig.\,\ref{Fig: MDPrecursor}.~The released Fe(g) atoms migrate within the simulation box until they are captured by an \ce{O2}(g) molecule, forming a \ce{FeO2}(g) radical, as shown in Fig.\,\ref{Fig: MDPrecursor}(a).~Subsequently, some \ce{FeO2}(g) radicals capture additional Fe(g) atoms, forming \ce{Fe2O2}, which then splits into FeO(g), as depicted in Fig.\,\ref{Fig: MDPrecursor}(b-d).

\begin{figure*}[h!]
\centering
\includegraphics[width=90mm]{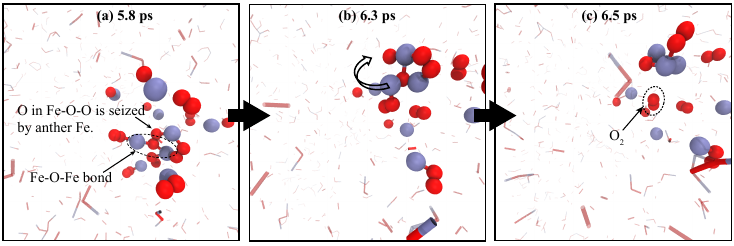}
   \caption{MD simulation (Phase II): aggregation reactions of \ce{FeO2}(g) into iron-oxide nascent nanocluster.~Atom colors: Fe (ice blue), O (red).}
\label{Fig: MDAggregation}
\end{figure*}

In Phase II of the MD investigations, our ReaxFF-MD trajectories reveal the evolution of iron-oxide nanoparticles originating from \ce{FeO2}(g) precursors.~As illustrated in Fig.\,\ref{Fig: MDAggregation}(a), adjacent \ce{FeO2} radicals are drawn together through Coulombic interactions between the negatively charged oxygen and the positively charged iron atoms.~This interaction leads to the formation of bridging Fe-O-Fe bonds within the emerging iron-oxide nanoclusters.~During this process, an oxygen atom, captured by another iron atom, breaks away from the O-O bond in the \ce{FeO2}(g) molecule.~This transferred oxygen atom also attracts additional radicals, facilitating further growth of the iron-oxide nanoclusters (Fig. \ref{Fig: MDAggregation}(b)).~During aggregation, a dangling \ce{O2} molecule detaches from the nanocluster, as depicted in Fig. \ref{Fig: MDAggregation}(c).

\begin{figure}[h!]
\centering
\includegraphics[width=75mm]{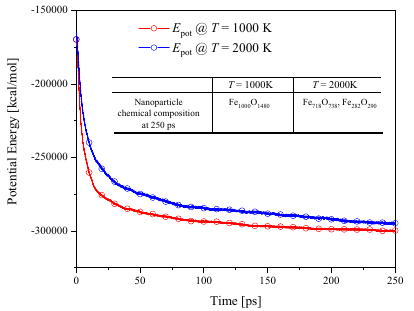}
   \caption{The potential energy of both temperatures reach a stable level by 250\,ps.}
\label{Fig: PotentialEnergy}
\end{figure}

Moreover, a comparative analysis of the chemical compositions of nanoparticles is conducted at 1000\,K and 2000\,K.~As illustrated in Fig.\,\ref{Fig: PotentialEnergy}, the potential energies at both temperatures stabilize over time, indicating that the chemical compositions of the iron-oxide nanoparticles reach a steady state by 250\,ps.~Specifically, at 1000\,K, all \ce{FeO2}(g) radicals agglomerate into a single nanoparticle with an atomic composition of \ce{Fe1000O1480}, corresponding to hematite (\ce{Fe2O3}).~In contrast, at 2000\,K, we observed the formation of two nanoparticles with chemical formulas of \ce{Fe718O738} and \ce{Fe282O290}, respectively.~This corresponds to a composition of FeO. Considering that the temperatures in the boundary layer of a diffusion-limited combusting particle are usually above 2000\,K, NPs are likely to consist of liquid FeO in the initial state.~This variation in chemical composition highlights the necessity for an ambient temperature below $\sim$2000\,K to form Fe(III)-dominant nanoparticles; at higher temperatures, Fe(II)-oxides are thermodynamically more stable than Fe(III)-oxides.~This agrees with the equilibrium phase diagram of the Fe-O system, suggesting that \ce{Fe2O3} is only stable below \SI{1740}{K}.

\section{Conclusions}
\label{sec:conclusions}

In this study, we investigated the critical processes of nanoparticle formation during the combustion of single iron particles using spatiotemporally-resolved, multi-physics, \textit{in-situ} optical diagnostics.~We experimentally evaluated key process parameters such as the initial particle size, nanoparticle initiation time and temperature, and the growth and evolution of nanoparticle clouds for spherical iron particles burning in three high-temperature atmospheres containing 20\%, 30\%, and 40\% oxygen by volume.~The growth and evolution of these nanoparticle clouds were further interpreted in conjunction with detailed CFD simulations.~Additionally, we explored the formation and aggregation of precursors, as well as the chemical composition of nanoparticle clusters, through ReaxFF-MD simulations.~The main conclusions of our study are summarized as follows.

\begin{enumerate}

\item The current diagnostics demonstrate a powerful capability to simultaneously measure particle size, nanoparticle initiation, and particle temperature for individual micrometer-sized iron particle as oxidation reactions progress.~However, the presence of NPs inevitably influences surface temperature measurements, particularly in high-oxygen environments.~Caution is advised when interpreting data, especially near peak temperatures, where NP clouds can significantly alter apparent emissivity.

\item In experiments, NP initiation time $t_\textrm{nano}$ is delayed for larger particles or with less oxygen.~This is because both scenarios require a longer particle residence time to reach the critical temperature for evaporation and NP formation.~However, NP initiation temperature $T_\textrm{prt, nano}$ shows only a weak dependency on particle diameter between 40 and 60\,\textmu m, while displaying a strong correlation with ambient oxygen mole fraction.

\item Detailed CFD simulations resolving the boundary layer processes near the particle surface suggest that the net velocity (convection + thermophoresis) around the particle can be either negative (toward the surface) or positive (away from the surface).~Negative transport could lead to NP attachment to microparticles, rendering them undetectable by line-of-sight measurements.~Consequently, visible NP clouds appear at a later time or higher temperature than the actual NP formation has started.~Exploring this process strongly supplements experiments to understand the higher observed $T_\textrm{prt, nano}$ at elevated oxygen concentrations.

\item MD simulations provide a detailed view of the fundamental processes involved in precursor formation and aggregation.~As Fe(g) atoms dissociate from the liquid phase, \ce{FeO2}(g) emerges as the first precursor, which further reacts to form \ce{FeO}(g).~By adopting the \ce{FeO2}(g) precursor, \ce{Fe2O3} nano clusters form at lower temperatures (e.g., 1000\,K), while FeO nanoclusters form at higher temperatures (e.g., 2000\,K).~Thus, precursor temperature is shown to be an important factor in determining the chemical composition of the resulting nanoclusters.

\end{enumerate}

\section*{Acknowledgement}
This work was founded by the Hessian Ministry of Higher Education, Research, Science and the Arts - cluster project Clean Circles.


%% References



%% New version of the num-names style
\bibliographystyle{elsarticle-num-names}
\bibliography{2024_Ref_Nanoparticles}

\begin{thebibliography}{56}
\expandafter\ifx\csname natexlab\endcsname\relax\def\natexlab#1{#1}\fi
\providecommand{\url}[1]{\texttt{#1}}
\providecommand{\href}[2]{#2}
\providecommand{\path}[1]{#1}
\providecommand{\DOIprefix}{doi:}
\providecommand{\ArXivprefix}{arXiv:}
\providecommand{\URLprefix}{URL: }
\providecommand{\Pubmedprefix}{pmid:}
\providecommand{\doi}[1]{\href{http://dx.doi.org/#1}{\path{#1}}}
\providecommand{\Pubmed}[1]{\href{pmid:#1}{\path{#1}}}
\providecommand{\bibinfo}[2]{#2}
\ifx\xfnm\relax \def\xfnm[#1]{\unskip,\space#1}\fi
%Type = Misc
\bibitem[{IEA(2024)}]{IEA.2024}
\bibinfo{author}{IEA}, \bibinfo{title}{Renewables 2023}, \bibinfo{year}{2024}. \URLprefix \url{https://www.iea.org/reports/renewables-2023}.
%Type = Misc
\bibitem[{IEA(2022)}]{IEA.2022}
\bibinfo{author}{IEA}, \bibinfo{title}{Iea energy and carbon tracker 2022}, \bibinfo{year}{2022}. \URLprefix \url{https://www.iea.org/data-and-statistics/data-product/iea-energy-and-carbon-tracker-2022}.
%Type = Article
\bibitem[{Neumann et~al.(2024)Neumann, Fradet, Scholtissek, Dammel, Riedel, Dreizler, Hasse, and Stephan}]{Neumann.2024}
\bibinfo{author}{J.~Neumann}, \bibinfo{author}{Q.~Fradet}, \bibinfo{author}{A.~Scholtissek}, \bibinfo{author}{F.~Dammel}, \bibinfo{author}{U.~Riedel}, \bibinfo{author}{A.~Dreizler}, \bibinfo{author}{C.~Hasse}, \bibinfo{author}{P.~Stephan},
\newblock \bibinfo{title}{Thermodynamic assessment of an iron-based circular energy economy for carbon-free power supply},
\newblock \bibinfo{journal}{Applied Energy} \bibinfo{volume}{368} (\bibinfo{year}{2024}) \bibinfo{pages}{123476}.
%Type = Article
\bibitem[{Bergthorson et~al.(2015)Bergthorson, Goroshin, Soo, Julien, Palecka, Frost, and Jarvis}]{Bergthorson.2015.Appl.Energy}
\bibinfo{author}{J.~M. Bergthorson}, \bibinfo{author}{S.~Goroshin}, \bibinfo{author}{M.~J. Soo}, \bibinfo{author}{P.~Julien}, \bibinfo{author}{J.~Palecka}, \bibinfo{author}{D.~L. Frost}, \bibinfo{author}{D.~J. Jarvis},
\newblock \bibinfo{title}{Direct combustion of recyclable metal fuels for zero-carbon heat and power},
\newblock \bibinfo{journal}{Applied Energy} \bibinfo{volume}{160} (\bibinfo{year}{2015}) \bibinfo{pages}{368--382}.
%Type = Article
\bibitem[{Bergthorson(2018)}]{Bergthorson.2018.Prog.EnergyCombust.Sci.}
\bibinfo{author}{J.~M. Bergthorson},
\newblock \bibinfo{title}{Recyclable metal fuels for clean and compact zero-carbon power},
\newblock \bibinfo{journal}{Progress in Energy and Combustion Science} \bibinfo{volume}{68} (\bibinfo{year}{2018}) \bibinfo{pages}{169--196}.
%Type = Article
\bibitem[{Dreizler et~al.(2021)Dreizler, Pitsch, Scherer, Schulz, and Janicka}]{Dreizler.2021.ApplEnergyCombustSci}
\bibinfo{author}{A.~Dreizler}, \bibinfo{author}{H.~Pitsch}, \bibinfo{author}{V.~Scherer}, \bibinfo{author}{C.~Schulz}, \bibinfo{author}{J.~Janicka},
\newblock \bibinfo{title}{The role of combustion science and technology in low and zero impact energy transformation processes},
\newblock \bibinfo{journal}{Applications in Energy and Combustion Science} \bibinfo{volume}{7} (\bibinfo{year}{2021}) \bibinfo{pages}{100040}.
%Type = Article
\bibitem[{Mi et~al.(2022)Mi, Fujinawa, and Bergthorson}]{Mi.2022}
\bibinfo{author}{X.~Mi}, \bibinfo{author}{A.~Fujinawa}, \bibinfo{author}{J.~M. Bergthorson},
\newblock \bibinfo{title}{A quantitative analysis of the ignition characteristics of fine iron particles},
\newblock \bibinfo{journal}{Combustion and Flame} \bibinfo{volume}{240} (\bibinfo{year}{2022}) \bibinfo{pages}{112011}.
%Type = Article
\bibitem[{Fujinawa et~al.(2023)Fujinawa, Thijs, Jean-Philyppe, Panahi, {Di Chang}, Schiemann, Levendis, Bergthorson, and Mi}]{Fujinawa.2023}
\bibinfo{author}{A.~Fujinawa}, \bibinfo{author}{L.~C. Thijs}, \bibinfo{author}{J.~Jean-Philyppe}, \bibinfo{author}{A.~Panahi}, \bibinfo{author}{{Di Chang}}, \bibinfo{author}{M.~Schiemann}, \bibinfo{author}{Y.~A. Levendis}, \bibinfo{author}{J.~M. Bergthorson}, \bibinfo{author}{X.~Mi},
\newblock \bibinfo{title}{Combustion behavior of single iron particles, part ii: A theoretical analysis based on a zero-dimensional model},
\newblock \bibinfo{journal}{Applications in Energy and Combustion Science} \bibinfo{volume}{14} (\bibinfo{year}{2023}) \bibinfo{pages}{100145}.
%Type = Article
\bibitem[{Nguyen et~al.(2024)Nguyen, Braig, Scholtissek, Ning, Li, Dreizler, and Hasse}]{Nguyen.2024}
\bibinfo{author}{B.-D. Nguyen}, \bibinfo{author}{D.~Braig}, \bibinfo{author}{A.~Scholtissek}, \bibinfo{author}{D.~Ning}, \bibinfo{author}{T.~Li}, \bibinfo{author}{A.~Dreizler}, \bibinfo{author}{C.~Hasse},
\newblock \bibinfo{title}{Ignition and kinetic-limited oxidation analysis of single iron microparticles in hot laminar flows},
\newblock \bibinfo{journal}{Fuel} \bibinfo{volume}{371} (\bibinfo{year}{2024}) \bibinfo{pages}{131866}.
%Type = Article
\bibitem[{Mich et~al.(2024)Mich, {Da Silva}, Ning, Li, Raabe, B{\"o}hm, Dreizler, Hasse, and Scholtissek}]{Mich.2024}
\bibinfo{author}{J.~Mich}, \bibinfo{author}{A.~K. {Da Silva}}, \bibinfo{author}{D.~Ning}, \bibinfo{author}{T.~Li}, \bibinfo{author}{D.~Raabe}, \bibinfo{author}{B.~B{\"o}hm}, \bibinfo{author}{A.~Dreizler}, \bibinfo{author}{C.~Hasse}, \bibinfo{author}{A.~Scholtissek},
\newblock \bibinfo{title}{Modeling the oxidation of iron microparticles during the reactive cooling phase},
\newblock \bibinfo{journal}{Proceedings of the Combustion Institute} \bibinfo{volume}{40} (\bibinfo{year}{2024}) \bibinfo{pages}{105538}.
%Type = Article
\bibitem[{Ning et~al.(2022)Ning, Shoshin, {van Oijen}, Finotello, and {de Goey}}]{Ning.2022.Combust.Flame}
\bibinfo{author}{D.~Ning}, \bibinfo{author}{Y.~Shoshin}, \bibinfo{author}{J.~A. {van Oijen}}, \bibinfo{author}{G.~Finotello}, \bibinfo{author}{L.~P. {de Goey}},
\newblock \bibinfo{title}{Critical temperature for nanoparticle cloud formation during combustion of single micron-sized iron particle},
\newblock \bibinfo{journal}{Combustion and Flame} \bibinfo{volume}{244} (\bibinfo{year}{2022}).
%Type = Article
\bibitem[{Panahi et~al.(2022)Panahi, {Di Chang}, Schiemann, Fujinawa, Mi, Bergthorson, and Levendis}]{Panahi.2022.ApplEnergyCombustSci}
\bibinfo{author}{A.~Panahi}, \bibinfo{author}{{Di Chang}}, \bibinfo{author}{M.~Schiemann}, \bibinfo{author}{A.~Fujinawa}, \bibinfo{author}{X.~Mi}, \bibinfo{author}{J.~M. Bergthorson}, \bibinfo{author}{Y.~A. Levendis},
\newblock \bibinfo{title}{Combustion behavior of single iron particles--part i: An experimental study in a drop-tube furnace under high heating rates and high temperatures},
\newblock \bibinfo{journal}{Applications in Energy and Combustion Science}  (\bibinfo{year}{2022}) \bibinfo{pages}{100097}.
%Type = Article
\bibitem[{Li et~al.(2022)Li, Heck, Reinauer, B{\"o}hm, and Dreizler}]{Li.2022.Combust.Flameb}
\bibinfo{author}{T.~Li}, \bibinfo{author}{F.~Heck}, \bibinfo{author}{F.~Reinauer}, \bibinfo{author}{B.~B{\"o}hm}, \bibinfo{author}{A.~Dreizler},
\newblock \bibinfo{title}{Visualizing particle melting and nanoparticle formation during single iron particle oxidation with multi-parameter optical diagnostics},
\newblock \bibinfo{journal}{Combustion and Flame} \bibinfo{volume}{245} (\bibinfo{year}{2022}) \bibinfo{pages}{112357}.
%Type = Article
\bibitem[{Ning et~al.(2023)Ning, Li, Mich, Scholtissek, B{\"o}hm, and Dreizler}]{Ning.2023}
\bibinfo{author}{D.~Ning}, \bibinfo{author}{T.~Li}, \bibinfo{author}{J.~Mich}, \bibinfo{author}{A.~Scholtissek}, \bibinfo{author}{B.~B{\"o}hm}, \bibinfo{author}{A.~Dreizler},
\newblock \bibinfo{title}{Multi-stage oxidation of iron particles in a flame-generated hot laminar flow},
\newblock \bibinfo{journal}{Combustion and Flame} \bibinfo{volume}{256} (\bibinfo{year}{2023}) \bibinfo{pages}{112950}.
%Type = Article
\bibitem[{Hameete et~al.(2024)Hameete, Abdallah, Thijs, Homan, Mi, Dam, and de~Goey}]{Hameete.2024}
\bibinfo{author}{J.~Hameete}, \bibinfo{author}{M.~S. Abdallah}, \bibinfo{author}{L.~C. Thijs}, \bibinfo{author}{T.~Homan}, \bibinfo{author}{X.~C. Mi}, \bibinfo{author}{N.~J. Dam}, \bibinfo{author}{L.~de~Goey},
\newblock \bibinfo{title}{Particle-resolved hyperspectral pyrometry of metal particles},
\newblock \bibinfo{journal}{Combustion and Flame} \bibinfo{volume}{264} (\bibinfo{year}{2024}) \bibinfo{pages}{113435}.
%Type = Article
\bibitem[{Tang et~al.(2011)Tang, Goroshin, and Higgins}]{Tang.2011}
\bibinfo{author}{F.-D. Tang}, \bibinfo{author}{S.~Goroshin}, \bibinfo{author}{A.~J. Higgins},
\newblock \bibinfo{title}{Modes of particle combustion in iron dust flames},
\newblock \bibinfo{journal}{Proceedings of the Combustion Institute} \bibinfo{volume}{33} (\bibinfo{year}{2011}) \bibinfo{pages}{1975--1982}.
%Type = Article
\bibitem[{McRae et~al.(2019)McRae, Julien, Salvo, Goroshin, Frost, and Bergthorson}]{McRae.2019}
\bibinfo{author}{M.~McRae}, \bibinfo{author}{P.~Julien}, \bibinfo{author}{S.~Salvo}, \bibinfo{author}{S.~Goroshin}, \bibinfo{author}{D.~L. Frost}, \bibinfo{author}{J.~M. Bergthorson},
\newblock \bibinfo{title}{Stabilized, flat iron flames on a hot counterflow burner},
\newblock \bibinfo{journal}{Proceedings of the Combustion Institute} \bibinfo{volume}{37} (\bibinfo{year}{2019}) \bibinfo{pages}{3185--3191}.
%Type = Article
\bibitem[{Fedoryk et~al.(2023)Fedoryk, Stelzner, Harth, and Trimis}]{Fedoryk.2023.ApplEnergyCombustSci}
\bibinfo{author}{M.~Fedoryk}, \bibinfo{author}{B.~Stelzner}, \bibinfo{author}{S.~Harth}, \bibinfo{author}{D.~Trimis},
\newblock \bibinfo{title}{Experimental investigation of the laminar burning velocity of iron-air flames in a tube burner},
\newblock \bibinfo{journal}{Applications in Energy and Combustion Science} \bibinfo{volume}{13} (\bibinfo{year}{2023}) \bibinfo{pages}{100111}.
%Type = Article
\bibitem[{Krenn et~al.(2024)Krenn, Li, Hebel, B{\"o}hm, and Dreizler}]{Krenn.2024}
\bibinfo{author}{T.~Krenn}, \bibinfo{author}{T.~Li}, \bibinfo{author}{J.~Hebel}, \bibinfo{author}{B.~B{\"o}hm}, \bibinfo{author}{A.~Dreizler},
\newblock \bibinfo{title}{Evaluation of a novel measurement method for the laminar burning speed in laminar lifted iron dust flames},
\newblock \bibinfo{journal}{Fuel} \bibinfo{volume}{366} (\bibinfo{year}{2024}) \bibinfo{pages}{131266}.
%Type = Article
\bibitem[{Goroshin et~al.(2022)Goroshin, Pale{\v{c}}ka, and Bergthorson}]{Goroshin.2022.Prog.EnergyCombust.Sci.}
\bibinfo{author}{S.~Goroshin}, \bibinfo{author}{J.~Pale{\v{c}}ka}, \bibinfo{author}{J.~M. Bergthorson},
\newblock \bibinfo{title}{Some fundamental aspects of laminar flames in nonvolatile solid fuel suspensions},
\newblock \bibinfo{journal}{Progress in Energy and Combustion Science} \bibinfo{volume}{91} (\bibinfo{year}{2022}) \bibinfo{pages}{100994}.
%Type = Article
\bibitem[{Li et~al.(2021)Li, Sanned, Huang, Berrocal, Cai, Ald{\'e}n, Richter, and Li}]{Li.2021.Opt.Express}
\bibinfo{author}{S.~Li}, \bibinfo{author}{D.~Sanned}, \bibinfo{author}{J.~Huang}, \bibinfo{author}{E.~Berrocal}, \bibinfo{author}{W.~Cai}, \bibinfo{author}{M.~Ald{\'e}n}, \bibinfo{author}{M.~Richter}, \bibinfo{author}{Z.~Li},
\newblock \bibinfo{title}{Stereoscopic high-speed imaging of iron microexplosions and nanoparticle-release},
\newblock \bibinfo{journal}{Optics express} \bibinfo{volume}{29} (\bibinfo{year}{2021}) \bibinfo{pages}{34465--34476}.
%Type = Article
\bibitem[{Li et~al.(2022)Li, Huang, Weng, Qian, Lu, Ald{\'e}n, and Li}]{Li.2022.Combust.Flame}
\bibinfo{author}{S.~Li}, \bibinfo{author}{J.~Huang}, \bibinfo{author}{W.~Weng}, \bibinfo{author}{Y.~Qian}, \bibinfo{author}{X.~Lu}, \bibinfo{author}{M.~Ald{\'e}n}, \bibinfo{author}{Z.~Li},
\newblock \bibinfo{title}{Ignition and combustion behavior of single micron-sized iron particle in hot gas flow},
\newblock \bibinfo{journal}{Combustion and Flame} \bibinfo{volume}{241} (\bibinfo{year}{2022}).
%Type = Article
\bibitem[{Cen et~al.(2024)Cen, Lyu, Qian, Li, and Lu}]{Cen.2024}
\bibinfo{author}{L.~Cen}, \bibinfo{author}{Z.~Lyu}, \bibinfo{author}{Y.~Qian}, \bibinfo{author}{Z.~Li}, \bibinfo{author}{X.~Lu},
\newblock \bibinfo{title}{In-situ light extinction nano-oxide volume fraction measurements during single iron particle combustion},
\newblock \bibinfo{journal}{Proceedings of the Combustion Institute} \bibinfo{volume}{40} (\bibinfo{year}{2024}) \bibinfo{pages}{105305}.
%Type = Article
\bibitem[{Ald{\'e}n(2022)}]{Alden.2022.Proc.Combust.Inst.}
\bibinfo{author}{M.~Ald{\'e}n},
\newblock \bibinfo{title}{Spatially and temporally resolved laser/optical diagnostics of combustion processes: From fundamentals to practical applications},
\newblock \bibinfo{journal}{Proceedings of the Combustion Institute}  (\bibinfo{year}{2022}).
%Type = Article
\bibitem[{Li et~al.(2024)Li, Farmand, Chen, Boehme, Nicolai, Hasse, Pitsch, and B{\"o}hm}]{Li.2024}
\bibinfo{author}{T.~Li}, \bibinfo{author}{P.~Farmand}, \bibinfo{author}{H.~Chen}, \bibinfo{author}{C.~Boehme}, \bibinfo{author}{H.~Nicolai}, \bibinfo{author}{C.~Hasse}, \bibinfo{author}{H.~Pitsch}, \bibinfo{author}{B.~B{\"o}hm},
\newblock \bibinfo{title}{Homogeneous ignition and volatile flame structure of single bituminous coal and walnut shell particles: Effects of particle size and gas atmosphere},
\newblock \bibinfo{journal}{Fuel} \bibinfo{volume}{371} (\bibinfo{year}{2024}) \bibinfo{pages}{131955}.
%Type = Article
\bibitem[{Ning et~al.(2024)Ning, Li, Li, B{\"o}hm, and Dreizler}]{Ning.2024}
\bibinfo{author}{D.~Ning}, \bibinfo{author}{Y.~Li}, \bibinfo{author}{T.~Li}, \bibinfo{author}{B.~B{\"o}hm}, \bibinfo{author}{A.~Dreizler},
\newblock \bibinfo{title}{Size-resolved ignition temperatures of isolated iron microparticles // temperature of burning iron microparticles with in-situ resolved initial sizes},
\newblock \bibinfo{journal}{Combustion and Flame} \bibinfo{volume}{270} (\bibinfo{year}{2024}) \bibinfo{pages}{113779}.
%Type = Article
\bibitem[{Abdallah et~al.(2024)Abdallah, Shoshin, Finotello, and de~Goey}]{Abdallah.2024}
\bibinfo{author}{M.~Abdallah}, \bibinfo{author}{Y.~Shoshin}, \bibinfo{author}{G.~Finotello}, \bibinfo{author}{L.~de~Goey},
\newblock \bibinfo{title}{Iron particles ignition in different hot coflow temperatures},
\newblock \bibinfo{journal}{Proceedings of the Combustion Institute} \bibinfo{volume}{40} (\bibinfo{year}{2024}) \bibinfo{pages}{105261}.
%Type = Article
\bibitem[{Ning et~al.(2022)Ning, Shoshin, {van Stiphout}, {van Oijen}, Finotello, and de~Goey}]{Ning.2022.Combust.Flameb}
\bibinfo{author}{D.~Ning}, \bibinfo{author}{Y.~Shoshin}, \bibinfo{author}{M.~{van Stiphout}}, \bibinfo{author}{J.~{van Oijen}}, \bibinfo{author}{G.~Finotello}, \bibinfo{author}{P.~de~Goey},
\newblock \bibinfo{title}{Temperature and phase transitions of laser-ignited single iron particle},
\newblock \bibinfo{journal}{Combustion and Flame} \bibinfo{volume}{236} (\bibinfo{year}{2022}).
%Type = Article
\bibitem[{Ning et~al.(2023)Ning, Shoshin, {van Oijen}, Finotello, and de~Goey}]{Ning.2022.Proc.Combust.Inst.}
\bibinfo{author}{D.~Ning}, \bibinfo{author}{Y.~Shoshin}, \bibinfo{author}{J.~{van Oijen}}, \bibinfo{author}{G.~Finotello}, \bibinfo{author}{P.~de~Goey},
\newblock \bibinfo{title}{Size evolution during laser-ignited single iron particle combustion},
\newblock \bibinfo{journal}{Proceedings of the Combustion Institute} \bibinfo{volume}{39} (\bibinfo{year}{2023}) \bibinfo{pages}{3561--3571}.
%Type = Article
\bibitem[{Huang et~al.(2022)Huang, Li, Sanned, Xu, Xu, Wang, Stiti, Qian, Cai, Berrocal, Richter, Ald{\'e}n, and Li}]{Huang.2022.Combust.Flame}
\bibinfo{author}{J.~Huang}, \bibinfo{author}{S.~Li}, \bibinfo{author}{D.~Sanned}, \bibinfo{author}{L.~Xu}, \bibinfo{author}{S.~Xu}, \bibinfo{author}{Q.~Wang}, \bibinfo{author}{M.~Stiti}, \bibinfo{author}{Y.~Qian}, \bibinfo{author}{W.~Cai}, \bibinfo{author}{E.~Berrocal}, \bibinfo{author}{M.~Richter}, \bibinfo{author}{M.~Ald{\'e}n}, \bibinfo{author}{Z.~Li},
\newblock \bibinfo{title}{A detailed study on the micro-explosion of burning iron particles in hot oxidizing environments},
\newblock \bibinfo{journal}{Combustion and Flame} \bibinfo{volume}{238} (\bibinfo{year}{2022}).
%Type = Article
\bibitem[{Thijs et~al.(2023)Thijs, {van Gool}, Ramaekers, {van Oijen}, and de~Goey}]{Thijs.2023}
\bibinfo{author}{L.~C. Thijs}, \bibinfo{author}{C.~{van Gool}}, \bibinfo{author}{W.~Ramaekers}, \bibinfo{author}{J.~A. {van Oijen}}, \bibinfo{author}{L.~de~Goey},
\newblock \bibinfo{title}{Resolved simulations of single iron particle combustion and the release of nano-particles},
\newblock \bibinfo{journal}{Proceedings of the Combustion Institute} \bibinfo{volume}{39} (\bibinfo{year}{2023}) \bibinfo{pages}{3551--3559}.
%Type = Article
\bibitem[{Thijs et~al.(2024)Thijs, Ning, Shoshin, Hazenberg, Mi, {van Oijen}, and de~Goey}]{Thijs.2024}
\bibinfo{author}{L.~C. Thijs}, \bibinfo{author}{D.~Ning}, \bibinfo{author}{Y.~S. Shoshin}, \bibinfo{author}{T.~Hazenberg}, \bibinfo{author}{X.~Mi}, \bibinfo{author}{J.~A. {van Oijen}}, \bibinfo{author}{P.~de~Goey},
\newblock \bibinfo{title}{Temperature evolution of laser-ignited micrometric iron particles: A comprehensive experimental data set and numerical assessment of laser heating impact},
\newblock \bibinfo{journal}{Applications in Energy and Combustion Science} \bibinfo{volume}{19} (\bibinfo{year}{2024}) \bibinfo{pages}{100284}.
%Type = Article
\bibitem[{Li et~al.(2023)Li, Geschwindner, Dreizler, and B{\"o}hm}]{Li.2023}
\bibinfo{author}{T.~Li}, \bibinfo{author}{C.~Geschwindner}, \bibinfo{author}{A.~Dreizler}, \bibinfo{author}{B.~B{\"o}hm},
\newblock \bibinfo{title}{Particle-resolved optical diagnostics of solid fuel combustion for clean power generation: a review},
\newblock \bibinfo{journal}{Measurement Science and Technology} \bibinfo{volume}{34} (\bibinfo{year}{2023}) \bibinfo{pages}{122001}.
%Type = Article
\bibitem[{T{\'o}th et~al.(2020)T{\'o}th, {\"O}gren, Sepman, Gren, and Wiinikka}]{Toth.2020.PowderTechnol.}
\bibinfo{author}{P.~T{\'o}th}, \bibinfo{author}{Y.~{\"O}gren}, \bibinfo{author}{A.~Sepman}, \bibinfo{author}{P.~Gren}, \bibinfo{author}{H.~Wiinikka},
\newblock \bibinfo{title}{Combustion behavior of pulverized sponge iron as a recyclable electrofuel},
\newblock \bibinfo{journal}{Powder Technology} \bibinfo{volume}{373} (\bibinfo{year}{2020}).
%Type = Article
\bibitem[{Poletaev and Khlebnikova(2020)}]{Poletaev.2020.Combust.Sci.Technol.}
\bibinfo{author}{N.~I. Poletaev}, \bibinfo{author}{M.~Y. Khlebnikova},
\newblock \bibinfo{title}{Combustion of iron particles suspension in laminar premixed and diffusion flames},
\newblock \bibinfo{journal}{Combustion Science and Technology}  (\bibinfo{year}{2020}) \bibinfo{pages}{1--22}.
%Type = Article
\bibitem[{Li et~al.(2021)Li, Farmand, Geschwindner, Greifenstein, K{\"o}ser, Schumann, Attili, Pitsch, Dreizler, and B{\"o}hm}]{Li.2021.Fuel}
\bibinfo{author}{T.~Li}, \bibinfo{author}{P.~Farmand}, \bibinfo{author}{C.~Geschwindner}, \bibinfo{author}{M.~Greifenstein}, \bibinfo{author}{J.~K{\"o}ser}, \bibinfo{author}{C.~Schumann}, \bibinfo{author}{A.~Attili}, \bibinfo{author}{H.~Pitsch}, \bibinfo{author}{A.~Dreizler}, \bibinfo{author}{B.~B{\"o}hm},
\newblock \bibinfo{title}{Homogeneous ignition and volatile combustion of single solid fuel particles in air and oxy-fuel conditions},
\newblock \bibinfo{journal}{Fuel} \bibinfo{volume}{291} (\bibinfo{year}{2021}) \bibinfo{pages}{120101}.
%Type = Misc
\bibitem[{Goodwin et~al.(2021)Goodwin, Speth, Moffat, and Weber}]{Goodwin.2021}
\bibinfo{author}{D.~G. Goodwin}, \bibinfo{author}{R.~L. Speth}, \bibinfo{author}{H.~K. Moffat}, \bibinfo{author}{B.~W. Weber}, \bibinfo{title}{Cantera: An object-oriented software toolkit for chemical kinetics, thermodynamics, and transport processes}, \bibinfo{year}{2021}. \DOIprefix\doi{10.5281/zenodo.4527812}.
%Type = Article
\bibitem[{K{\"o}ser et~al.(2019)K{\"o}ser, Li, Vorobiev, Dreizler, Schiemann, and B{\"o}hm}]{Koser.2019.Proc.Combust.Inst.}
\bibinfo{author}{J.~K{\"o}ser}, \bibinfo{author}{T.~Li}, \bibinfo{author}{N.~Vorobiev}, \bibinfo{author}{A.~Dreizler}, \bibinfo{author}{M.~Schiemann}, \bibinfo{author}{B.~B{\"o}hm},
\newblock \bibinfo{title}{Multi-parameter diagnostics for high-resolution in-situ measurements of single coal particle combustion},
\newblock \bibinfo{journal}{Proceedings of the Combustion Institute} \bibinfo{volume}{37} (\bibinfo{year}{2019}) \bibinfo{pages}{2893--2900}.
%Type = Article
\bibitem[{Li et~al.(2021)Li, Schiemann, K{\"o}ser, Dreizler, and B{\"o}hm}]{Li.2021.RenewableSustainableEnergyRev.}
\bibinfo{author}{T.~Li}, \bibinfo{author}{M.~Schiemann}, \bibinfo{author}{J.~K{\"o}ser}, \bibinfo{author}{A.~Dreizler}, \bibinfo{author}{B.~B{\"o}hm},
\newblock \bibinfo{title}{Experimental investigations of single particle and particle group combustion in a laminar flow reactor using simultaneous volumetric oh-lif imaging and diffuse backlight-illumination},
\newblock \bibinfo{journal}{Renewable and Sustainable Energy Reviews} \bibinfo{volume}{136} (\bibinfo{year}{2021}) \bibinfo{pages}{110377}.
%Type = Article
\bibitem[{Li et~al.(2023)Li, Li, Farmand, Dreizler, Pitsch, and B{\"o}hm}]{Li.2022.Proc.Combust.Inst.b}
\bibinfo{author}{T.~Li}, \bibinfo{author}{B.~Li}, \bibinfo{author}{P.~Farmand}, \bibinfo{author}{A.~Dreizler}, \bibinfo{author}{H.~Pitsch}, \bibinfo{author}{B.~B{\"o}hm},
\newblock \bibinfo{title}{Motion and swelling of single coal particles during volatile combustion in a laminar flow reactor},
\newblock \bibinfo{journal}{Proceedings of the Combustion Institute} \bibinfo{volume}{39} (\bibinfo{year}{2023}) \bibinfo{pages}{3333--3341}.
%Type = Article
\bibitem[{Levendis et~al.(1992)Levendis, Estrada, and Hottel}]{Levendis.1992.Rev.Sci.Instrum}
\bibinfo{author}{Y.~A. Levendis}, \bibinfo{author}{K.~R. Estrada}, \bibinfo{author}{H.~C. Hottel},
\newblock \bibinfo{title}{Development of multicolor pyrometers to monitor the transient response of burning carbonaceous particles},
\newblock \bibinfo{journal}{The Review of scientific instruments} \bibinfo{volume}{63} (\bibinfo{year}{1992}) \bibinfo{pages}{3608--3622}.
%Type = Article
\bibitem[{Nguyen et~al.(2024)Nguyen, Scholtissek, Li, Ning, Dreizler, and Hasse}]{nguyen2024}
\bibinfo{author}{B.-D. Nguyen}, \bibinfo{author}{A.~Scholtissek}, \bibinfo{author}{T.~Li}, \bibinfo{author}{D.~Ning}, \bibinfo{author}{A.~Dreizler}, \bibinfo{author}{C.~Hasse},
\newblock \bibinfo{title}{Nanoparticle formation in the boundary layer of burning iron microparticles: modeling and simulation},
\newblock \bibinfo{journal}{submitted to Chemical Engineering Journal}  (\bibinfo{year}{2024}).
%Type = Article
\bibitem[{Friedlander(2000)}]{friendlander2000}
\bibinfo{author}{S.~K. Friedlander},
\newblock \bibinfo{title}{Smoke, dust, and haze: Fundamentals of aerosol dynamics}  (\bibinfo{year}{2000}).
%Type = Article
\bibitem[{Finke and Sewerin(2024)}]{finke2024}
\bibinfo{author}{J.~Finke}, \bibinfo{author}{F.~Sewerin},
\newblock \bibinfo{title}{A population balance approach for predicting the size distribution of oxide smoke near a burning aluminum particle},
\newblock \bibinfo{journal}{Combustion and Flame} \bibinfo{volume}{265} (\bibinfo{year}{2024}) \bibinfo{pages}{113464}.
%Type = Article
\bibitem[{Panda and Pratsinis(1995)}]{panda1995}
\bibinfo{author}{S.~Panda}, \bibinfo{author}{S.~E. Pratsinis},
\newblock \bibinfo{title}{Modeling the synthesis of aluminum particles by evaporation-condensation in an aerosol flow reactor},
\newblock \bibinfo{journal}{Nanostructured Materials} \bibinfo{volume}{5} (\bibinfo{year}{1995}) \bibinfo{pages}{755--767}.
%Type = Article
\bibitem[{Janbazi et~al.(2019)Janbazi, Karakaya, Kasper, Schulz, Wlokas, and Peukert}]{janbazi2019}
\bibinfo{author}{H.~Janbazi}, \bibinfo{author}{Y.~Karakaya}, \bibinfo{author}{T.~Kasper}, \bibinfo{author}{C.~Schulz}, \bibinfo{author}{I.~Wlokas}, \bibinfo{author}{S.~Peukert},
\newblock \bibinfo{title}{Development and evaluation of a chemical kinetics reaction mechanism for tetramethylsilane-doped flames},
\newblock \bibinfo{journal}{Chemical Engineering Science} \bibinfo{volume}{209} (\bibinfo{year}{2019}) \bibinfo{pages}{115209}.
%Type = Article
\bibitem[{{van Duin} et~al.(2001){van Duin}, Dasgupta, Lorant, and Goddard}]{vanDuin.2001}
\bibinfo{author}{A.~C.~T. {van Duin}}, \bibinfo{author}{S.~Dasgupta}, \bibinfo{author}{F.~Lorant}, \bibinfo{author}{W.~A. Goddard},
\newblock \bibinfo{title}{Reaxff: A reactive force field for hydrocarbons},
\newblock \bibinfo{journal}{The Journal of Physical Chemistry A} \bibinfo{volume}{105} (\bibinfo{year}{2001}) \bibinfo{pages}{9396--9409}.
%Type = Article
\bibitem[{Senftle et~al.(2016)Senftle, Hong, Islam, Kylasa, Zheng, Shin, Junkermeier, Engel-Herbert, Janik, Aktulga, Verstraelen, Grama, and {van Duin}}]{Senftle.2016}
\bibinfo{author}{T.~P. Senftle}, \bibinfo{author}{S.~Hong}, \bibinfo{author}{M.~M. Islam}, \bibinfo{author}{S.~B. Kylasa}, \bibinfo{author}{Y.~Zheng}, \bibinfo{author}{Y.~K. Shin}, \bibinfo{author}{C.~Junkermeier}, \bibinfo{author}{R.~Engel-Herbert}, \bibinfo{author}{M.~J. Janik}, \bibinfo{author}{H.~M. Aktulga}, \bibinfo{author}{T.~Verstraelen}, \bibinfo{author}{A.~Grama}, \bibinfo{author}{A.~C.~T. {van Duin}},
\newblock \bibinfo{title}{The reaxff reactive force-field: development, applications and future directions},
\newblock \bibinfo{journal}{npj Computational Materials} \bibinfo{volume}{2} (\bibinfo{year}{2016}).
%Type = Inproceedings
\bibitem[{Uene et~al.(2019)Uene, Mabuchi, Zaitsu, Yasuhara, and Tokumasu}]{Uene.2019}
\bibinfo{author}{N.~Uene}, \bibinfo{author}{T.~Mabuchi}, \bibinfo{author}{M.~Zaitsu}, \bibinfo{author}{S.~Yasuhara}, \bibinfo{author}{T.~Tokumasu},
\newblock \bibinfo{title}{Molecular dyanamics simulation of thermal chemical vapor deposition for hydrogenated amorphous silicon on si (100) substrate by reactive force-field},
\newblock in: \bibinfo{booktitle}{2019 International Conference on Simulation of Semiconductor Processes and Devices (SISPAD)}, \bibinfo{publisher}{IEEE}, \bibinfo{year}{2019}, pp. \bibinfo{pages}{1--4}. \DOIprefix\doi{10.1109/SISPAD.2019.8870438}.
%Type = Article
\bibitem[{Shin et~al.(2015)Shin, Kwak, Vasenkov, Sengupta, and {van Duin}}]{Shin.2015}
\bibinfo{author}{Y.~K. Shin}, \bibinfo{author}{H.~Kwak}, \bibinfo{author}{A.~V. Vasenkov}, \bibinfo{author}{D.~Sengupta}, \bibinfo{author}{A.~C. {van Duin}},
\newblock \bibinfo{title}{Development of a reaxff reactive force field for fe/cr/o/s and application to oxidation of butane over a pyrite-covered cr 2 o 3 catalyst},
\newblock \bibinfo{journal}{ACS Catalysis} \bibinfo{volume}{5} (\bibinfo{year}{2015}) \bibinfo{pages}{7226--7236}.
%Type = Article
\bibitem[{Rumminger et~al.(1999)Rumminger, Reinelt, Babushok, and Linteris}]{Rumminger.1999}
\bibinfo{author}{M.~D. Rumminger}, \bibinfo{author}{D.~Reinelt}, \bibinfo{author}{V.~Babushok}, \bibinfo{author}{G.~T. Linteris},
\newblock \bibinfo{title}{Numerical study of the inhibition of premixed and diffusion flames by iron pentacarbonyl11official contribution of the national institute of standards and technology; not subject to copyright in the united states},
\newblock \bibinfo{journal}{Combustion and Flame} \bibinfo{volume}{116} (\bibinfo{year}{1999}) \bibinfo{pages}{207--219}.
%Type = Article
\bibitem[{Nanjaiah et~al.(2021)Nanjaiah, Pilipodi-Best, Lalanne, Fjodorow, Schulz, Cheskis, Kempf, Wlokas, and Rahinov}]{Nanjaiah.2021}
\bibinfo{author}{M.~Nanjaiah}, \bibinfo{author}{A.~Pilipodi-Best}, \bibinfo{author}{M.~R. Lalanne}, \bibinfo{author}{P.~Fjodorow}, \bibinfo{author}{C.~Schulz}, \bibinfo{author}{S.~Cheskis}, \bibinfo{author}{A.~Kempf}, \bibinfo{author}{I.~Wlokas}, \bibinfo{author}{I.~Rahinov},
\newblock \bibinfo{title}{Experimental and numerical investigation of iron-doped flames: Feo formation and impact on flame temperature},
\newblock \bibinfo{journal}{Proceedings of the Combustion Institute} \bibinfo{volume}{38} (\bibinfo{year}{2021}) \bibinfo{pages}{1249--1257}.
%Type = Article
\bibitem[{Gao et~al.(2022)Gao, Zhu, Wang, Yilmaz, and {van Duin}}]{Gao.2022}
\bibinfo{author}{Y.~Gao}, \bibinfo{author}{W.~Zhu}, \bibinfo{author}{T.~Wang}, \bibinfo{author}{D.~E. Yilmaz}, \bibinfo{author}{A.~C.~T. {van Duin}},
\newblock \bibinfo{title}{C/h/o/f/al reaxff force field development and application to study the condensed-phase poly(vinylidene fluoride) and reaction mechanisms with aluminum},
\newblock \bibinfo{journal}{The Journal of Physical Chemistry C} \bibinfo{volume}{126} (\bibinfo{year}{2022}) \bibinfo{pages}{11058--11074}.
%Type = Article
\bibitem[{Sperling et~al.(2025)Sperling, Deutschmann, Ning, Spielmann, Li, Kramm, Nirschl, B{\"o}hm, and Dreizler}]{Sperling.2025}
\bibinfo{author}{A.~Sperling}, \bibinfo{author}{M.~P. Deutschmann}, \bibinfo{author}{D.~Ning}, \bibinfo{author}{J.~Spielmann}, \bibinfo{author}{T.~Li}, \bibinfo{author}{U.~I. Kramm}, \bibinfo{author}{H.~Nirschl}, \bibinfo{author}{B.~B{\"o}hm}, \bibinfo{author}{A.~Dreizler},
\newblock \bibinfo{title}{Oxidation progress and inner structure during single micron-sized iron particles combustion in a hot oxidizing atmosphere},
\newblock \bibinfo{journal}{Fuel} \bibinfo{volume}{381} (\bibinfo{year}{2025}) \bibinfo{pages}{133147}.
%Type = Article
\bibitem[{{Joseph Kalman} et~al.(2015){Joseph Kalman}, {Nick Glumac}, {and Herman Krier}, Kalman, Glumac, and Krier}]{JosephKalman.2015}
\bibinfo{author}{{Joseph Kalman}}, \bibinfo{author}{{Nick Glumac}}, \bibinfo{author}{{and Herman Krier}}, \bibinfo{author}{J.~Kalman}, \bibinfo{author}{N.~Glumac}, \bibinfo{author}{H.~Krier},
\newblock \bibinfo{title}{High-temperature metal oxide spectral emissivities for pyrometry applications},
\newblock \bibinfo{journal}{Journal of Thermophysics and Heat Transfer} \bibinfo{volume}{29} (\bibinfo{year}{2015}) \bibinfo{pages}{874--879}.
%Type = Article
\bibitem[{Yetter et~al.(2009)Yetter, Risha, and Son}]{Yetter.2009.Proc.Combust.Inst.}
\bibinfo{author}{R.~A. Yetter}, \bibinfo{author}{G.~A. Risha}, \bibinfo{author}{S.~F. Son},
\newblock \bibinfo{title}{Metal particle combustion and nanotechnology},
\newblock \bibinfo{journal}{Proceedings of the Combustion Institute} \bibinfo{volume}{32} (\bibinfo{year}{2009}) \bibinfo{pages}{1819--1838}.

\end{thebibliography}


\begin{thebibliography}{1}
\expandafter\ifx\csname natexlab\endcsname\relax\def\natexlab#1{#1}\fi
\providecommand{\url}[1]{\texttt{#1}}
\providecommand{\href}[2]{#2}
\providecommand{\path}[1]{#1}
\providecommand{\DOIprefix}{doi:}
\providecommand{\ArXivprefix}{arXiv:}
\providecommand{\URLprefix}{URL: }
\providecommand{\Pubmedprefix}{pmid:}
\providecommand{\doi}[1]{\href{http://dx.doi.org/#1}{\path{#1}}}
\providecommand{\Pubmed}[1]{\href{pmid:#1}{\path{#1}}}
\providecommand{\bibinfo}[2]{#2}
\ifx\xfnm\relax \def\xfnm[#1]{\unskip,\space#1}\fi
%Type = Article
\bibitem[{Ning et~al.(2022)Ning, Shoshin, {van Stiphout}, {van Oijen}, Finotello, and de~Goey}]{Ning.2022.Combust.Flameb}
\bibinfo{author}{D.~Ning}, \bibinfo{author}{Y.~Shoshin}, \bibinfo{author}{M.~{van Stiphout}}, \bibinfo{author}{J.~{van Oijen}}, \bibinfo{author}{G.~Finotello}, \bibinfo{author}{P.~de~Goey},
\newblock \bibinfo{title}{Temperature and phase transitions of laser-ignited single iron particle},
\newblock \bibinfo{journal}{Combustion and Flame} \bibinfo{volume}{236} (\bibinfo{year}{2022}).

\end{thebibliography}






% \end{thebibliography}

\end{document}

\maketitle

\section*{CFD Simulation Validation}
\renewcommand{\thefigure}{S1}

The particle model used in the CFD simulation is validated by comparing results to measurements of laser-ignited particles by Ning~et~al.~\cite{Ning.2022.Combust.Flameb}. The boundary conditions are $T_\mathrm{g}=\SI{300}{\kelvin}$, $X_{\ce{O2}}=0.21$, $X_{\ce{N2}}=0.79$, $p=\SI{1e5}{\pascal}$, $d_\mathrm{p,0}=\SI{50}{\micro\meter}$, and $T_\mathrm{p,0}=\SI{1500}{\kelvin}$. Figure~\ref{Fig:validation} shows the temporal evolution of the particle temperature. The numerical results are in a good agreement with the experimental measurements. The peak temperature in the simulation is around \SI{2482}{\kelvin} at $t=\SI{25.4}{\milli\second}$, while it is around \SI{2429}{\kelvin} at $t=\SI{24}{\milli\second}$ in the experiment.

\renewcommand{\thefigure}{S1}
\begin{figure}[h!]
\centering
    \includegraphics[width=75mm]{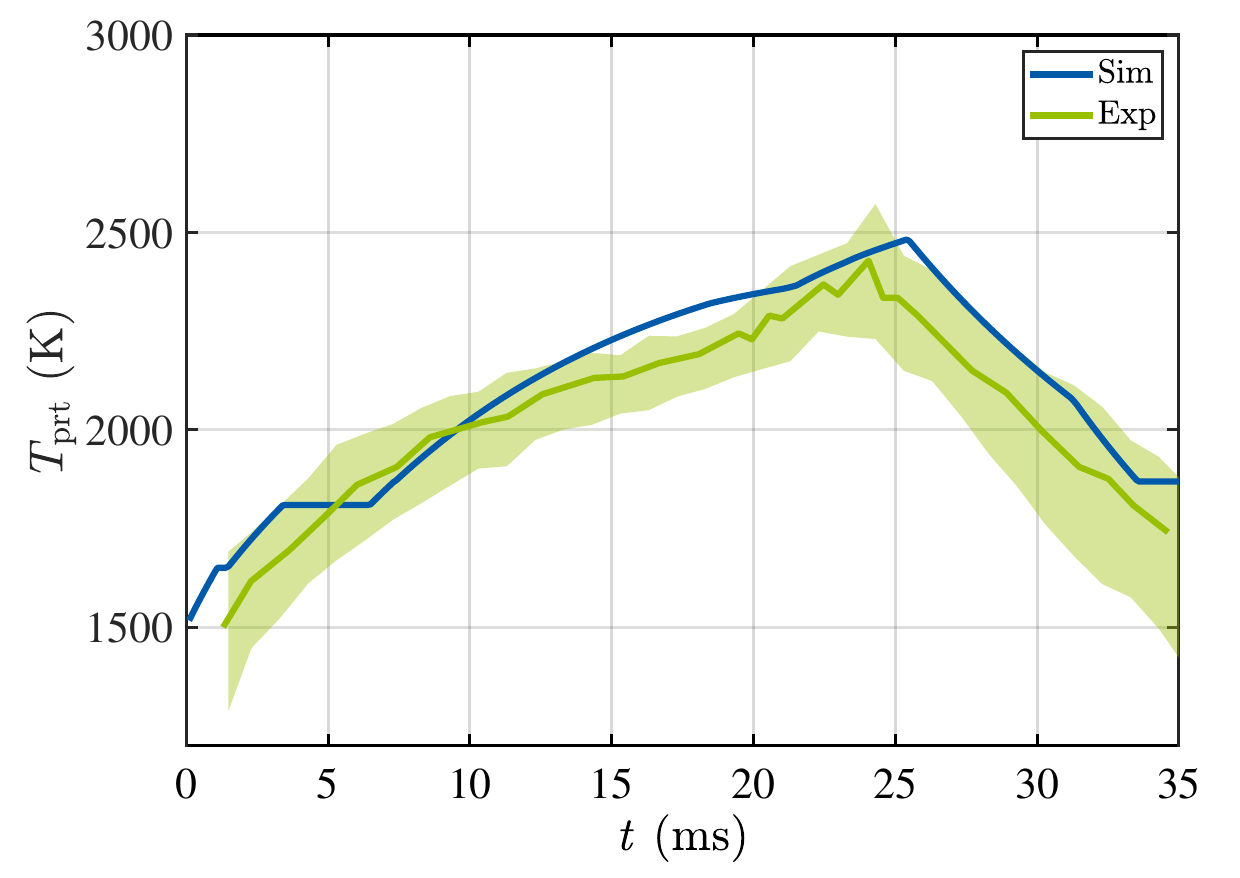}
   \caption{CFD Simulation: Numerically obtained particle temperature evolution compared with experimental measurements by Ning~et~al.~\cite{Ning.2022.Combust.Flameb}.}
\label{Fig:validation}
\end{figure}

\section*{Evolution of nanoparticle volume}

Figure \ref{Fig_Sim_NPVolume_over_time} depicts the temporal evolution of $\Sigma V_\textrm{nano, norm}$ for three particle sizes in 20vol\% (AIR20), 30vol\% (AIR30), and 40vol\% (AIR40) atmospheres.~Numerical times $t_\mathrm{nano}$ are determined at the location of maximum curvature of $\Sigma V_\textrm{nano, norm}$, as shown by the vertical dotted lines.~The trends of numerical simulations agree qualitatively with the experiments, in which the intensity integration $\Sigma I_\textrm{DBI,nano}$ represents the accumulation and growth of NP clouds.

\renewcommand{\thefigure}{S2}
\begin{figure}[h!]
\centering
\includegraphics[width=75mm]{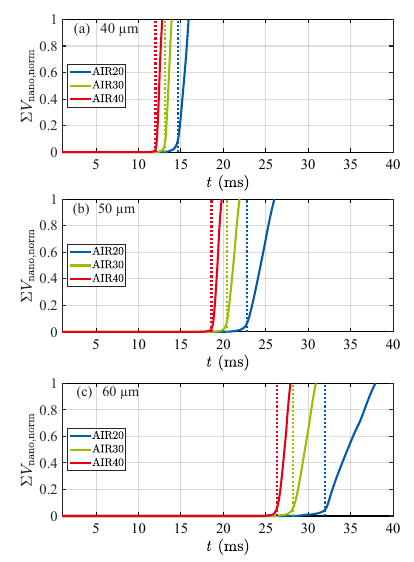}
   \caption{CFD Simulation: normalized nanoparticle volume integration $\Sigma V_\textrm{nano, norm}$ (solid lines) and determined nanoparticle initiation time (dashed lines) at maximum curvature of the solid lines.}
\label{Fig_Sim_NPVolume_over_time}
\end{figure}

\section*{MD-NVE simulation temperature profile}

Figure {Fig: MDTemperature} depicts the particle temporal evolution predicted by the molecule dynamics simulation.~In the Phase I, the initial temperature is set to 1500\,K, which increases rapidly to about 4500\,K.

\renewcommand{\thefigure}{S3}
\begin{figure}[h!]
\centering
\includegraphics[width=75mm]{Figures/Fig_MD_Temperature Profile.pdf}
   \caption{The temperature profile during the MD-NVE simulation.~Atom colors: Fe (ice blue), and O (red).}
\label{Fig: MDTemperature}
\end{figure}

\bibliographystyle{elsarticle-num-names}
\bibliography{2024_Ref_Nanoparticles}